\documentclass[twocolumn,twoside]{IEEEtran}
\usepackage{mathpple}
\usepackage{times}
\usepackage{amsmath}  
\usepackage{amssymb}  
\usepackage{mathrsfs} 
\usepackage{theorem}  
\usepackage{cite}     
\usepackage{comment}  
\usepackage{upref}
\usepackage{amsfonts}
\usepackage{verbatim}
\usepackage[dvipsnames,usenames]{color}
\usepackage{enumerate}
\usepackage{graphicx}
\usepackage{subfigure}
\usepackage{latexsym}

\usepackage{psfrag}
\usepackage{makecell}

\newcommand{\code}{\ttfamily\bfseries}

\newcommand{\be}[1]{\begin{equation}\label{#1}}
\newcommand{\ee}{\end{equation}}

\newcommand{\bc}{\begin{center}}
\newcommand{\ec}{\end{center}}

\newcommand{\floor}[1]{\lfloor{#1}\rfloor}

\newcommand{\cB}{{\cal B}}
\newcommand{\cC}{{\cal C}}

\newcommand{\cE}{{\cal E}}

\newcommand{\cG}{{\cal G}}
\newcommand{\cH}{{\cal H}}

\newcommand{\cT}{{\cal T}}

\newcommand{\cW}{{\cal W}}
\newcommand{\cX}{{\cal X}}

\newcommand{\bfb}{{\boldsymbol b}}

\newcommand{\bfe}{{\boldsymbol e}}

\newcommand{\bfi}{{\boldsymbol i}}
\newcommand{\bfj}{{\boldsymbol j}}

\newcommand{\bfw}{{\boldsymbol w}}
\newcommand{\bfx}{{\boldsymbol x}}
\newcommand{\bfy}{{\boldsymbol y}}
\newcommand{\bfz}{{\boldsymbol z}}

\newcommand{\bfI}{{\mathbf I}}

\renewcommand{\leq}{\leqslant}

\renewcommand{\geq}{\geqslant}

\newcommand{\Cref}[1]{Co\-rol\-la\-ry\,\ref{#1}}

\theoremstyle{plain} \theorembodyfont{\normalfont\slshape}

\newtheorem{thm}{Theorem$\!$}
\newenvironment{theorem}{\begin{thm}\hspace*{-1ex}{\bf.}}{\end{thm}}

\newtheorem{prop}[thm]{Proposition$\!$}

\newtheorem{lem}[thm]{Lemma$\!$}
\newenvironment{lemma}{\begin{lem}\hspace*{-1ex}{\bf.}}{\end{lem}}

\newtheorem{cor}[thm]{Corollary$\!$}
\newenvironment{corollary}{\begin{cor}\hspace*{-1ex}{\bf.}}{\end{cor}}

\newtheorem{defi}{Definition.}

\theorembodyfont{\normalfont}

\newtheorem{exam}{Example$\!$}
\newenvironment{example}{\begin{exam}\hspace*{-1ex}{\bf .}}{\end{exam}}

\newtheorem{remrk}{Remark$\!$}

\definecolor{Codecolor}{named}{White}  

\newcommand{\Copen}{\mbox{\{\kern-5.50pt\{}}
\newcommand{\Cclose}{\mbox{\}\kern-5.50pt\}}}
\newcommand{\Cslash}{\mbox{$\backslash\kern-6.02pt\backslash$}}

\makeatletter

\DeclareRobustCommand{\sbinom}{\genfrac[]\z@{}}

\makeatother

\newcommand{\G}[2]{\sbinom{{#1}\kern-.05pt}{{#2}\kern-.05pt}}

\newcommand{\Gq}[2]{\sbinom{{#1}}{{#2}}_2}

\newcommand{\Gb}[2]{\sbinom{{#1}}{{#2}}_{2}}
\begin{document}
\title{\textbf{\huge{Generalized Sphere Packing Bound}}}

\author{\textbf{Arman Fazeli},\!\IEEEauthorrefmark{1}
        \IEEEauthorblockN{\textbf{Alexander Vardy},\!\IEEEauthorrefmark{1}
        and \textbf{Eitan Yaakobi}\IEEEauthorrefmark{2}}

\IEEEauthorblockA{\IEEEauthorrefmark{1}University of California San Diego, La Jolla, CA 92093, USA \\}
\IEEEauthorblockA{\IEEEauthorrefmark{2}California Institute of Technology, Pasadena, CA 91125, USA\\}
{\code  \{afazelic,avardy\}@ucsd.edu},\,
\code  yaakobi@caltech.edu\vspace{-5ex}}\maketitle

\begin{abstract}

Kulkarni and Kiyavash recently introduced~a~new method
to establish upper bounds on the size of deletion-correct\-ing
codes. This 
method is based upon tools from hypergraph~theory.
The deletion channel is represented by a hypergraph whose\linebreak
edges are the deletion balls (or spheres), so
that a deletion-correct\-ing code becomes a matching
in this hypergraph. Consequently, a~bound on the
size of such a code can be obtained from bounds
on the matching number of a hypergraph. Classical
results in~hypergraph theory are then invoked 
to compute an upper bound on the matching number as a solution
to a linear-programming problem: the problem of finding fractional
transversals.

The method by Kulkarni and Kiyavash can be applied not only for the deletion channel but also for other error channels. This paper studies this method in its most general setup. First, it is shown that if the error channel is regular and symmetric then the upper bound by this method coincides with the well-known \emph{sphere packing bound} and thus is called here the \emph{generalized sphere packing bound}. Even though this bound is explicitly given by a linear programming problem, finding its exact value may still be a challenging task. The art of finding the exact upper bound (or slightly weaker ones) is the assignment of weights to the hypergraph's vertices in a way that they satisfy the constraints in the linear programming problem. In order to simplify the complexity of the linear programming, we present a technique based upon graph automorphisms that in many cases significantly reduces the number of variables and constraints in the problem. We then apply this method on specific examples of error channels. We start with the $Z$ channel and show how to exactly find the generalized sphere packing bound for this setup. Next studied is the non-binary limited magnitude channel both for symmetric and asymmetric errors, where we focus on the single-error case. We follow up on the deletion channel, which was the original motivation of the work by Kulkarni and Kiyavash, and show how to improve upon their upper bounds for single-deletion-correcting codes. Since the deletion and grain-error channels resemble a very similar structure for a single error, we also improve upon the existing upper bounds on single-grain error-correcting codes. Finally, we apply this method for projective spaces and find its generalized sphere packing bound for the single-error case.
\end{abstract}

\section{Introduction}\label{sec:intro}
One of the basic and fundamental results in coding theory asserts that an upper bound on a length-$n$ binary code $\cC$ with minimum Hamming distance
$2r+1$ is $$|\cC|\leq \frac{2^n}{B(r)},$$ where $B(r)=\sum_{i=0}^r\binom{n}{i}$. This is known as the classical \textbf{\emph{sphere packing bound}}.
This bound can be applied for other cases as well. Let $X$ be a finite set with some distance function $d:X\times X\rightarrow \mathbb{N}$.
Assume that the volume of every ball is the same, that is, if $B_r(x) \triangleq \{y\in X \ | \ d(x,y)\leq r\}$ then for all $x\in X$, $|B_r(x)| = \Delta_r$ for some fixed value $\Delta_r$. Then, the resulting sphere packing bound on a code $\cC\subseteq X$ with minimum distance $2r+1$ becomes $|X|/\Delta_r$. However, \textbf{\emph{what happens if the size of all balls is not the same?}} Clearly, a naive solution is to use $\Delta_r$ as the minimum size of all balls and then to apply the same bound, but this approach can give a very weak upper bound. The goal of this paper is to study a generalization of the sphere packing bound for setups where the size of all balls is not necessarily the same.

The lower counter bound for the sphere packing one is the well-known Gilbert-Varshamov bound~\cite{G52,V57}. This bound states that if the size of all balls of radius $r$ is the same, $\Delta_{r}$, then a lower bound on a code $\cC\subseteq X$ with minimum distance $r+1$ becomes $|X|/\Delta_{r}$. In~\cite{T97}, a similar study was carried for the Gilbert-Varshamov bound in case that the size of all balls is not necessarily the same. Using Tur\'{a}n's theorem, it was shown that the same derivation on a lower bound of a code still holds, with the modification of using the average size of the balls. That is, if $\overline{\Delta}_{r}\triangleq (\sum_{x\in X}|B_{r}(x)|)/|X|$, then a generalized Gilbert-Varshamov bound asserts that there exists a code with minimum distance $r+1$ and of size at least $|X|/\overline{\Delta}_{r}$. Thus, an immediate question to ask is whether the same analogy holds for the sphere packing bound: \textbf{\emph{Is $|X|/\overline{\Delta}_{r}$ an upper bound on a code $\cC\subseteq X$ with minimum distance $2r+1$?}} Even though in most of the cases we study in this work this derivation does hold, the answer in general to this question is negative. However, it is interesting to find some conditions under which this bound will always be satisfied.

The deletion channel~\cite{S62} is one of the examples where the balls can have different sizes. Recently, in~\cite{KK12}, Kulkarni and Kiyavash showed a technique, based upon tools from hypergraph theory~\cite{B79}, in order to derive explicit non-asymptotic upper bounds on the cardinalities of deletion-correcting codes. These upper bounds were given both for binary and non-binary codes as well as for deletion-correcting codes for constrained sources. Since the method in~\cite{KK12} can be applied for other similar setups, more results were presented shortly after for different channel models. Upper bounds on the cardinalities of grain-error-correcting codes were given in~\cite{GYD13} and~\cite{KZ13} and similar bounds for multipermutations codes with the Kendall's $\tau$ distance were derived in~\cite{BYEB13}.

This paper has two main goals. First, we extend the method studied for the deletion channel by Kulkarni and Kiyavash~\cite{KK12} and analyze it in its most general setting. We assume that the error channel is characterized by a directed graph, which depicts for a given transmitted word, its set of possible received words. Then, an upper bound will be given on codes which can correct $r$ errors, for some fixed $r$. This bound is established by the solution of a linear programming given from a hypergraph that is derived from the error channel graph. In particular, it is shown that the sphere packing bound is a special case of this bound. We also study properties of this bound and show a scheme, based upon graph automorphisms, that in many cases can significantly reduce the complexity of the linear programming problem. In the second part of this work, we provide specific examples on the application of this method to setups where the balls have different sizes. These examples include the $Z$ channel, non-binary channels with limited magnitude errors (symmetric and asymmetric), deletion channel, grain-error channel, and finally, projective spaces. In some of these examples we improve upon the existing results which use this method to calculate the upper bound on the code cardinalities. When possible in these examples, we compare the bounds we receive with the state-of-the-art ones.

In order to describe our results, we need to introduce some notation. Let $\cH=(X,\cE)$ be a hypergraph, where $X=\{x_1,\ldots, x_n\}$ is its vertices set and $\cE=\{E_1,\ldots, E_m\}$ is its hyperedges set. Let $A$ be the $n\times m$ incidence matrix of $\cH$, so $A(i,j)=1$ if $x_i\in E_j$. A \textbf{\emph{transversal}} in $\cH$ is a subset $T\subseteq X$ that intersects every hyperedge in $\cE$.  The \textbf{\emph{transversal number}} of $\cH$, denoted by $\tau(\cH)$, is the size of the smallest transversal. Every transversal can be represented by a binary vector $\bfw\in\{0,1\}^n$ which needs to satisfy $A^T\cdot \bfw \geq \textbf{1}$. However, if the vector $\bfw$ can have values over $\mathbb{R}_+$ and still satisfies the last inequality, then it is called a \textbf{\emph{fractional transversal}}. Under this setup, it is known that $\tau^*(\cH)\leq \tau(\cH)$, where $\tau^*(\cH)$ is the linear programming relaxation of $\tau(\cH)$, defined as
\begin{equation}\label{eq:GSPB}
\tau^*(\cH) = \min\bigg\{\sum_{i=1}^n w_i : A^T\cdot\bfw \geq \mathbf{1}, \bfw\in \mathbb{R}_+^n \bigg\}.
\end{equation}

Let $\cG=(X,E)$ be a directed graph which describes an error channel. The vertices set $X$ is the set of all possible transmitted words, and the edges set $E$ consists of all pairs of vertices of distance one. The distance between $x,y\in X$, is the path metric in $\cG$ and is denoted by $d(x,y)$. Note that since the graph is directed, it is possible to have $d(x,y)\neq d(y,x)$. For every $x\in X$, its radius-$r$ ball is the set $B_r(x)$ which was defined above and its degree is $\deg_r(x) = |B_r(x)|$. The largest cardinality of a length-$n$ code in $\cG$ with minimum distance $d$ is denoted by $A_{\cG}(n,d)$. Given some positive integer $r$, the graph $\cG$ is associated with a hypergraph $\cH(\cG,r)=(X_r,\cE_r)$ where $X_r=X$ and $\cE_r=\{B_r(x)\ | \ x\in X\}$. Observing that every code $\cC\subseteq X$ of minimum distance $2r+1$ is a matching in $\cH(\cG,r)$ (which is a collection of pairwise disjoint edges), the following upper bound on $A_{\cG}(n,2r+1)$ was verified in~\cite{KK12},
\begin{equation}\label{eq:bound}
A_{\cG}(n,2r+1) \leq \tau^*(\cH(\cG,r)).
\end{equation}

One of the first properties we present asserts that if the graph $\cG$ is regular such that $\deg_r(x)=\Delta_r$ for all $x\in X$, and the distance function $d$ is symmetric, then the bound $\tau^*(\cH(\cG,r))$ coincides with the sphere packing bound, that is, $\tau^*(\cH(\cG,r)) = \frac{|X|}{\Delta_r}$. Therefore, in this work the bound $\tau^*(\cH(\cG,r))$ is called the \textbf{\emph{generalized sphere packing bound}}.

The expression $\tau^*(\cH(\cG,r))$ provides an explicit upper bound on $A_{\cG}(n,2r+1)$. However, it may still be a hard problem to calculate this value since it requires the solution of a linear programming problem that can have an exponential number of variables and constraints. Clearly, one would inspire to find this exact value, but if this is not possible to accomplish, it is still valuable to give an upper bound on $\tau^*(\cH(\cG,r))$, which, in essence, is an upper bound on $A_{\cG}(n,2r+1)$ as well. Such an upper bound will be given by finding any fractional transversal and the goal will be to find one with small weight. In fact, all the upper bound results presented in~\cite{BYEB13,GYD13,KZ13,KK12} follow this approach and an upper bound on the value $\tau^*(\cH(\cG,r))$ in each case is given.

The rest of the paper is organized as follows. Section~\ref{sec:def} establishes the rest of the definitions and tools required in this paper and demonstrates them on the $Z$ channel. This channel will be used throughout the paper as a running example and a case study we rigorously investigate. In Section~\ref{sec:gen res}, we start with basic properties on the generalized sphere packing bound. In particular, we show upper and lower bounds on its value and prove that if the graph $\cG$ is regular and symmetric then the sphere packing bound coincides with the generalized sphere packing bound. We also show several examples which establish a dissenting answer to the question brought earlier about the upper bound validity of an average sphere packing value. We then proceed to define a special monotonicity property on the graph $\cG$ which states that a graph is monotone if for all $r$ and two vertices $x$ and $y$, if $y\in B_r(x)$ then $\deg_r(y) \leq \deg_r(x)$. This property is useful in order to give a general formula for a fractional transversal and a corresponding upper bound. In fact, this property and fractional transversal were used in the previous works~\cite{GYD13,KZ13,KK12}. Lastly in this section, we use tools from automorphisms on graphs in order to simplify the complexity of the linear programming problem in~(\ref{eq:GSPB}). Noticing that in many channels there are groups of vertices with similar behavior motivates us to treat them as the same vertex and thus significantly reduce the number of variables and constraints in the linear programming~(\ref{eq:GSPB}). In Section~\ref{sec:Z channel}, we study the $Z$ channel. Our main contribution here is finding a method to calculate the generalized sphere packing bound for all radii. In Section~\ref{sec:non binary} we carry a similar task for the limited-magnitude channel with symmetric and asymmetric errors. We focus only the single error case of radius one in both cases and find fractional transversals and corresponding upper bounds. Section~\ref{sec:deletions n grains} follows upon the original work of~\cite{KK12}, improving the bounds derived therein for the deletion channel (for the case of a single deletion). Since the structure of the deletion and grain-error channel is very similar, especially for a single error, we continue with the same approach to improve upon the existing upper bounds from~\cite{GYD13,KZ13} on the cardinalities of single-grain error-correcting codes. Section~\ref{sec:projective} studies bounds on projective spaces and in particular we give an optimal solution for the radius-one case under this channel. Finally, Section~\ref{sec:conclusion} concludes the paper and proposes some problems which remained open.

\section{Definitions and Preliminaries}\label{sec:def}
In this section we formally define the tools and definitions used throughout the paper. We mainly follow the same definitions and properties from~\cite{KK12}.

Let $\cH=(X,\cE)$ be a hypergraph where $X=\{x_1,\ldots, x_n\}$, $\cE=\{E_1,\ldots, E_m\}$ and $A$ its $n\times m$ incidence matrix. A \textbf{\emph{matching}} in $\cH$ is a collection of pairwise disjoint hyperedges and the \textbf{\emph{matching number}} of $\cH$, denoted by $\nu(\cH)$, is the size of the largest matching. The matching number of $\cH$, $\nu(\cH)$, is the solution of the integer linear programming problem
$$\nu(\cH)  = \max\bigg\{\sum_{i=1}^m z_i : A\cdot\bfz \leq \mathbf{1}, \bfz\in \{0,1\}^m \bigg\}.$$ Note that the transversal number $\tau(\cH)$, defined in the previous section, is the solution of the integer linear programming problem
$$\tau(\cH) = \min\bigg\{\sum_{i=1}^n w_i : A^T\cdot\bfw \geq \mathbf{1}, \bfw\in \{0,1\}^n \bigg\}.$$
These two problems satisfy weak duality and thus $\nu(\cH) \leq \tau(\cH)$. Furthermore, they can be slightly modified such that the vectors in the minimization and maximization problems can have values in $\mathbb{Z}_+$, and still they give the values of $\nu(\cH)$ and $\tau(\cH)$, that is,
\begin{align*}
& \nu(\cH)  = \max\bigg\{\sum_{i=1}^m z_i : A\cdot\bfz \leq \mathbf{1}, \bfz\in \mathbb{Z}_+^m \bigg\},& \\
& \tau(\cH) = \min\bigg\{\sum_{i=1}^n w_i : A^T\cdot\bfw \geq \mathbf{1}, \bfw\in \mathbb{Z}_+^n \bigg\}.
\end{align*}

The relaxation of these integer linear programmings allows the variables $\bfz$ and $\bfw$ to take values in $\mathbb{R}_+$, which are not necessarily integers. The value of this linear programming relaxation for the matching number is denoted by
$$ \nu^*(\cH)  = \max\bigg\{\sum_{i=1}^m z_i : A\cdot\bfz \leq \mathbf{1}, \bfz\in \mathbb{R}_+^m \bigg\},$$
and the corresponding one for the transversal number is the value $\tau^*(\cH)$, stated in~(\ref{eq:GSPB}).
Note that the real solutions can be significantly different than the integer solutions and since $\nu^*(\cH)$ and $\tau^*(\cH)$ satisfy strong duality, the following property holds~\cite{KK12} $$\nu(\cH) \leq \nu^*(\cH) = \tau^*(\cH) \leq \tau(\cH),$$
and in particular, for any fractional transversal $\bfw$, $$\nu(\cH) \leq \tau^*(\cH) \leq \sum_{i=1}^n w_i.$$
Lastly, we mention here that we will usually denote the fractional transversal by $\bfw=(w_1,\ldots,w_n)$, such that $w_i$ corresponds to the value that is assigned to the vertex $x_i$. However, when it will be clear from the context, the notation $w_x$ will be used to refer to the value of $w_i$, where $x=x_i$.

Every error channel studied in this work will be depicted by some directed graph $\cG=(X,E)$, where the set $E$ defines the set of all pairs of vertices of distance one from each other. The distance between every two vertices $x,y\in X$, denoted by $d(x,y)$, is the length of the shortest path from $x$ to $y$ in the graph $\cG$, and $d(x,y) = \infty$ if such a path does not exist. Note that this definition of distance is not necessarily symmetric and thus it may happen that $d(x,y)\neq d(y,x)$. However if for all $x,y\in X$, $d(x,y) = d(y,x)$, then we say that $\cG$ is \textbf{\emph{symmetric}}, and otherwise it is \textbf{\emph{not symmetric}}. For any $x\in X$, we let $B_r^\textmd{out}(x), B_r^\textmd{in}(x)$ be the sets $B_r^\textmd{out}(x)=\{y\in X \ | \ d(x,y)\leq r\}$ and $B_r^\textmd{in}(x)=\{y\in X \ | \ d(y,x)\leq r\}$. The out-degree of $x$ is $\deg_r^\textmd{out}(x) = |B_r^\textmd{out}(x)|$ and the in-degree is $\deg_r^\textmd{in}(x) = |B_r^\textmd{in}(x)|$. The definition of $B_r^\textmd{out}(x)$ and $\deg_r^\textmd{out}(x)$ coincide with the ones in the Introduction for $B_r(x)$ and $\deg_r(x)$, respectively. To ease the notation in the paper we will follow the ones from the Introduction for the ``out'' case and use the ones defined above for the ``in'' case.

If a word $x\in X$ is transmitted and at most $r$ errors occurred then any word in $B_r(x)$ can be received. A code $\cC\subseteq X$ in this graph is said to have minimum distance $d$ if for all $x,y\in \cC$, $d(x,y)\geq d$. We let $A_{\cG}(n,d)$ be the largest cardinality of a code in $\cG$ of length $n$ and minimum distance $d$. If for every $r\geq 0$, there exists some fixed $\Delta_r$ such that for every $x\in X$, $\deg_r(x)=\Delta_r$, then we say that the graph $\cG$ is \textbf{\emph{regular}} and otherwise it is called \textbf{\emph{non-regular}}.

For any positive integer $r$, $\cH(\cG,r)=(X_r,\cE_r)$ is a hypergraph associated with $\cG$ such that $X_r=X$ and $\cE_r=\{B_r(x) : x\in X\}$. As was stated in~({\ref{eq:bound}), the value $\tau^*(\cH(\cG,r))$ is an upper bound on $A_{\cG}(n,2r+1)$ and is called in this work the generalized sphere packing bound.

The average size of a ball of radius $r$ in $\cG$ is defined to be
$$\overline{\Delta}_r = \frac{1}{|X|}\sum_{x\in X}\deg_r(x).$$ In~\cite{T97}, using Tur\'{a}n's theorem a generalized Gilbert-Varshamov bound was shown to hold also for the cases where the size of all balls is not the same. This bound asserts that a lower bound on $A_{\cG}(n,d)$ is given by $$\frac{|X|}{\overline{\Delta}_{d-1}} \leq A_{\cG}(n,d).$$

Let us remind the question we brought in the Introduction about the analogy of the last bound to the sphere packing bound. Namely, does the following inequality hold $$ A_{\cG}(n,2r+1)\leq \frac{|X|}{\overline{\Delta}_r}?$$
We call the value $\frac{|X|}{\overline{\Delta}_r}$ the \textbf{\emph{average sphere packing value}} and denote it by $ASPV(\cG,r)$. We do not call this value a bound since, as we shall see later, it is not necessarily a valid upper bound.

The following example demonstrates the definitions and concepts introduced in this section for the $Z$ channel.
\begin{example}\label{ex:Z1}
The $Z$ channel is a channel with binary inputs and outputs where the errors are asymmetric. Here, we assume that errors can only change a 1 to 0 with some probability $0<p<1$, but not vice versa; see Fig~\ref{fig:Z channel}.
\begin{figure}%
\centering
\includegraphics[width =0.5\linewidth]{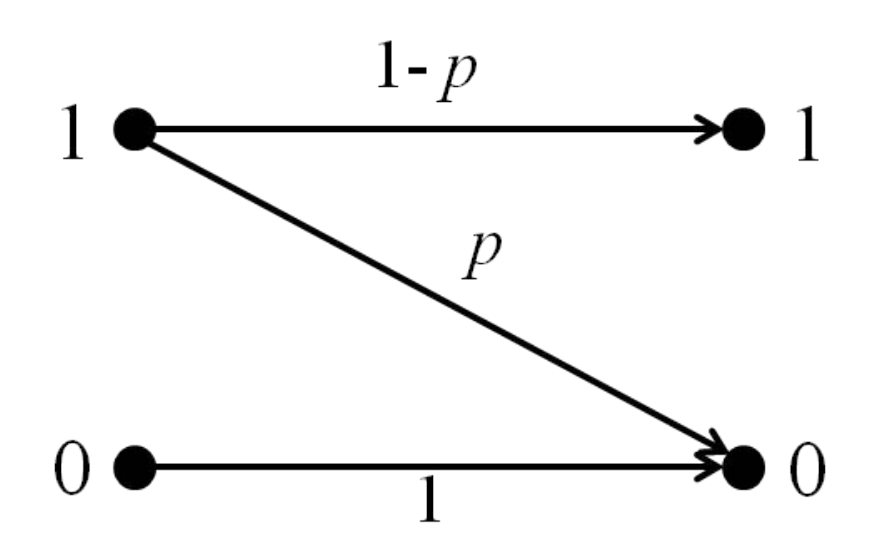}%
\caption{The $Z$-channel.}
\label{fig:Z channel}%
\end{figure}
The corresponding graph is $\cG_Z=(X_Z,E_Z)$, where $X_Z=\{0,1\}^n$ and
$$E_Z = \{ (\bfx,\bfy) : \bfx,\bfy\in \{0,1\}^n, \bfx\geq \bfy, w_H(\bfx) = w_H(\bfy)+1\},$$
and $w_H(\bfx)$ denotes the Hamming weight of $\bfx$. Let $r$ be some fixed positive integer. For every $\bfx\in \{0,1\}^n$, $$B_{Z,r}(\bfx) = \{\bfy \in \{0,1\}^n \ : \ \bfx\geq \bfy, w_H(\bfx)-w_H(\bfy)\leq r\},$$ and $\deg_{Z,r}(\bfx) = \sum_{i=0}^r\binom{w_H(\bfx)}{i}$.

The corresponding hypergraph is $\cH(\cG_Z,r) = (X_{Z,r},\cE_{Z,r})$, such that $X_{Z,r} = \{0,1\}^n$ and $\cE_{Z,r} = \{B_{Z,r}(\bfx) \ : \ \bfx\in \{0,1\}^n \}$. The generalized sphere packing bound becomes
\begin{small}
\begin{equation}\label{eq:Z GSPB}
\hspace{-1ex}\tau^*(\cH(\cG_Z,r))\hspace{-0.3ex}=\hspace{-0.3ex}\min \bigg\{\hspace{-0.8ex}\sum_{\bfx\in\{0,1\}^n}\hspace{-2ex}w_{\bfx} :\hspace{-0.5ex} \forall \bfx\in\hspace{-0.8ex}\{0,1\}^n,\hspace{-1.5ex} \sum_{\bfy\in B_{Z,r}(\bfx)}\hspace{-2ex}w_{\bfy} \geq 1, w_{\bfx}\geq 0  \bigg\}.
\end{equation}
\end{small}
The average size of a ball with radius $r$ is
\begin{align*}
\overline{\Delta}_{Z,r} & = \frac{1}{2^n}\sum_{\bfx\in\{0,1\}^n}\sum_{i=0}^r\binom{w_H(\bfx)}{i} =  \frac{1}{2^n}\sum_{w=0}^n\binom{n}{w}\sum_{i=0}^r\binom{w}{i} &\\
& = \frac{1}{2^n}\sum_{i=0}^r\sum_{w=0}^n\binom{n}{w}\binom{w}{i}.
\end{align*}
For $0\leq i\leq r$, $\sum_{w=0}^n\binom{n}{w}\binom{w}{i} = \binom{n}{i}2^{n-i}$ and thus we get
\begin{align*}
\overline{\Delta}_{Z,r} = \frac{1}{2^n}\sum_{i=0}^r\binom{n}{i}2^{n-i} = \sum_{i=0}^r\frac{\binom{n}{i}}{2^{i}} .
\end{align*}
Therefore, the average sphere packing value in this case becomes
$$ASPV(\cG_Z,r)=\frac{2^n}{\overline{\Delta}_{Z,r}} = \frac{2^n}{\sum_{i=0}^r\frac{\binom{n}{i}}{2^{i}}}.$$
In particular, for $r=1$ we get
$$ASPV(\cG_Z,1)=\frac{2^n}{\overline{\Delta}_{Z,1}} = \frac{2^n}{1+n/2} = \frac{2^{n+1}}{n+2}.$$
In the sequel it will be verified that the average sphere packing value for $r=1$ is a valid upper bound for the $Z$ channel.
\hfill$\Box$
\end{example}

Even though the generalized sphere packing bound $\tau^*(\cH(\cG,r))$ gives an explicit upper bound on the cardinality of error-correcting codes, it is not necessarily immediate to calculate it. To accomplish this task, one needs to solve a linear programming which, in general, does not necessarily have an efficient solution. Furthermore, note that in many of the communication channels the number of variables and constraints can be very large and in particular exponential with the length of the words. Our main discussion in this paper will be dedicated towards approaches for deriving the value $\tau^*(\cH(\cG,r))$ for different graphs $\cG$.  However, in cases where it will not be possible to derive this explicit value, we note that every fractional transversal provides a valid upper bound and thus we inspire to give the best fractional transversal we can find.

\section{General Results and Observations}\label{sec:gen res}

In this section we start by proving basic properties on the value of the generalized sphere packing bound $\tau^*(\cH(\cG,r))$ as specified in~(\ref{eq:GSPB}). We then show some approaches for finding fractional transversals. Finally, we present a scheme, based upon automorphisms on graphs, that in many cases can significantly reduce the complexity of the linear programming problem for calculating the value $\tau^*(\cH(\cG,r))$. As specified in Section~\ref{sec:def}, we assume throughout this section that the error channel is depicted by some directed graph $\cG=(X,E)$ and for a fixed integer $r\geq 1$, $\cH(\cG,r)=(X_r,\cE_r)$ is its associated hypergraph.

\subsection{Basic Properties of the Generalized Sphere Packing Bound}\label{subsec:basic properties}
We start here by proving some basic properties and giving insights on the value of $\tau^*(\cH(\cG,r))$.
The next lemma proves a lower bound on the generalized sphere packing bound in case that its in-degree is upper bounded.
\begin{lemma}\label{lem:lower}
If for all $x\in X$, $\deg_r^\textmd{in}(x)\leq \Delta$, then $$\tau^*(\cH(\cG,r)) \geq \frac{|X|}{\Delta}.$$
\end{lemma}
\begin{IEEEproof}
Since $\deg_r^\textmd{in}(x)\leq \Delta$, for all $x\in X$, the weight of every column of the incidence matrix $A$ of $\cH(\cG,r)$ is at most $\Delta$, that is, $\sum_{i=1}^na_{i,j}\leq \Delta$ for all $1\leq j\leq n$. Let $\bfw$ be a fractional transversal in $\cH(\cG,r)$. Then, for every $1\leq i\leq n$, $\sum_{j=1}^na_{i,j}w_j\geq 1$, and thus
$$n\leq \sum_{i=1}^n\sum_{j=1}^na_{i,j}w_j.$$
However, note that
$$\hspace{-0.5ex} n\leq \hspace{-0.5ex}\sum_{i=1}^n\sum_{j=1}^na_{i,j}w_j =\hspace{-0.5ex} \sum_{j=1}^n\sum_{i=1}^na_{i,j}w_j =\hspace{-0.5ex} \sum_{j=1}^nw_j\hspace{-0.5ex}\sum_{i=1}^na_{i,j} \leq\hspace{-0.5ex} \Delta\sum_{j=1}^nw_j,$$
and therefore $$\sum_{j=1}^nw_j \geq \frac{n}{\Delta}.$$
Hence, we conclude that $\tau^*(\cH(\cG,r)) \geq \frac{|X|}{\Delta}$.
\end{IEEEproof}
Next, we show an upper bound on the generalized sphere packing bound in case that its out-degree is lower bounded.
\begin{lemma}\label{lem:upper}
If for all $x\in X$, $\deg_r(x)\geq \Delta$, then
$$\tau^*(\cH(\cG,r)) \leq \frac{|X|}{\Delta}.$$
\end{lemma}
\begin{IEEEproof}
If $\deg_r(x)\geq \Delta$ for all $x\in X$ then the vector $\bfw= \textbf{1}/\Delta$ is a fractional transversal and thus $\tau^*(\cH(\cG,r)) \leq |X|/\Delta$.
\end{IEEEproof}

According to the last two lemmas we can show that if the graph $\cG$ is regular and symmetric then the generalized sphere packing bound coincides with the sphere packing bound.
\begin{corollary}
If the graph $\cG$ is symmetric and regular then the generalized sphere packing bound and the sphere packing bound coincide. Furthermore, $\tau^*(\cH(\cG,r)) = \frac{|X|}{\Delta_r}$, where for all $x\in X$, $\deg_r(x) = \deg_r^{\textmd{in}}(x)=\Delta_r$.
\end{corollary}
\begin{IEEEproof}
Since $\cG$ is regular then for all $x\in X$, $\det_r(x) = \Delta_r$ and according to Lemma~\ref{lem:upper}, we have $\tau^*(\cH(\cG,r)) \leq \frac{|X|}{\Delta_r}$. Since $\cG$ is also symmetric we have that for all $x\in X$, $\deg_r^{\textmd{in}}(x) = \Delta_r$ and according to Lemma~\ref{lem:lower}, we get $\tau^*(\cH(\cG,r)) \geq \frac{|X|}{\Delta_r}$. Therefore, $\tau^*(\cH(\cG,r)) = \frac{|X|}{\Delta_r}$.
\end{IEEEproof}

The next example proves that the requirement on the graph $\cG$ to be symmetric is necessary in order to have equality between the sphere packing and the generalized sphere packing bound.
\begin{example}\label{ex:symmetric prop}
In this example the graph $\cG_2=(X_2,E_2)$ has six vertices, so $X_2=\{x_1,x_2,x_3,x_4,x_5,x_6\}$. For $2\leq i\leq 6$, there is an edge from $x_i$ to $x_1$ and finally there is an edge from $x_1$ to $x_2$; see Fig.~\ref{fig:ex sym}.
\begin{figure}%
\centering
\includegraphics[width =0.7\linewidth]{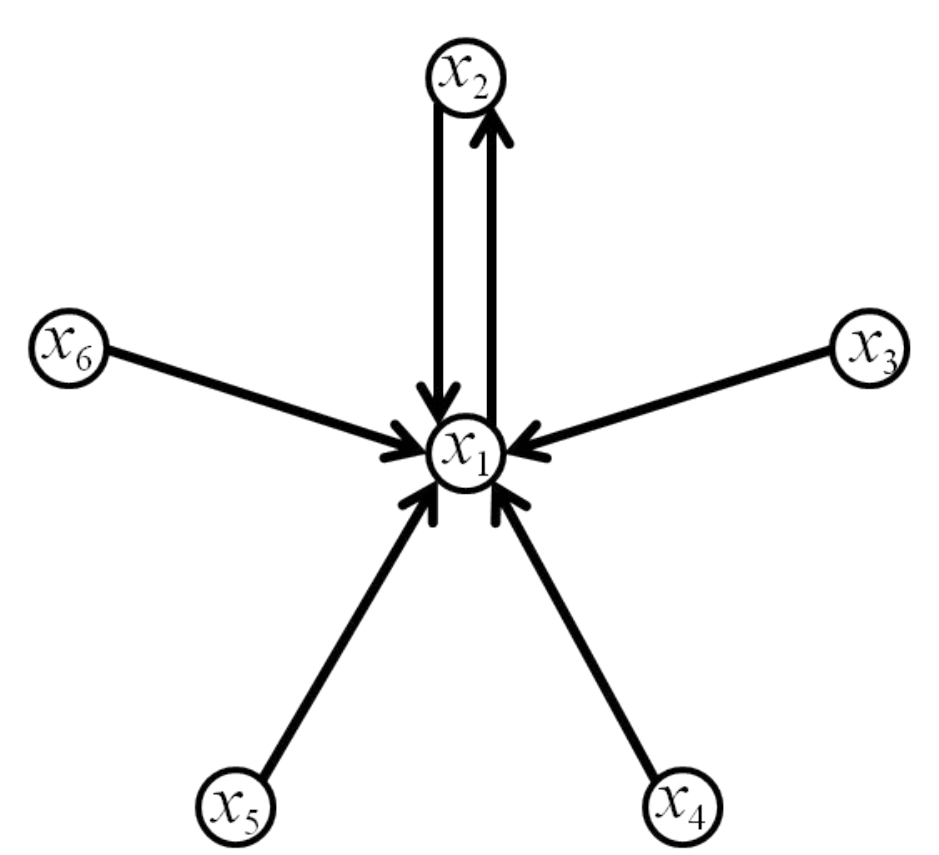}%
\caption{The graph $\cG_2$.}
\label{fig:ex sym}%
\end{figure}
Therefore, $b_1(x_i) = 2$ for all $1\leq i\leq 6$, so the graph $\cG_2$ is regular and the sphere packing bound becomes $\frac{|X_2|}{2}=3$. However, the vector $\bfw = (1,0,0,0,0,0)$ is a fractional transversal, which is optimal, and thus the generalized sphere packing bound of $\cG_2$ equals 1.
\hfill$\Box$
\end{example}

In the next example, we show a graph that does not obey to the average sphere packing value. This provides a negative answer to the earlier question we asked in the Introduction regarding the validity of the average sphere packing value as a valid bound.
\begin{example}\label{ex3}
The graph $\cG_3=(X_3,E_3)$ in this example has five vertices, so $X_3=\{x_1,x_2,x_3,x_4,x_5\}$. There is an edge from the first vertex to all other four vertices; see Fig.~\ref{fig:ex3}.
\begin{figure}%
\centering
\includegraphics[width =0.5\linewidth]{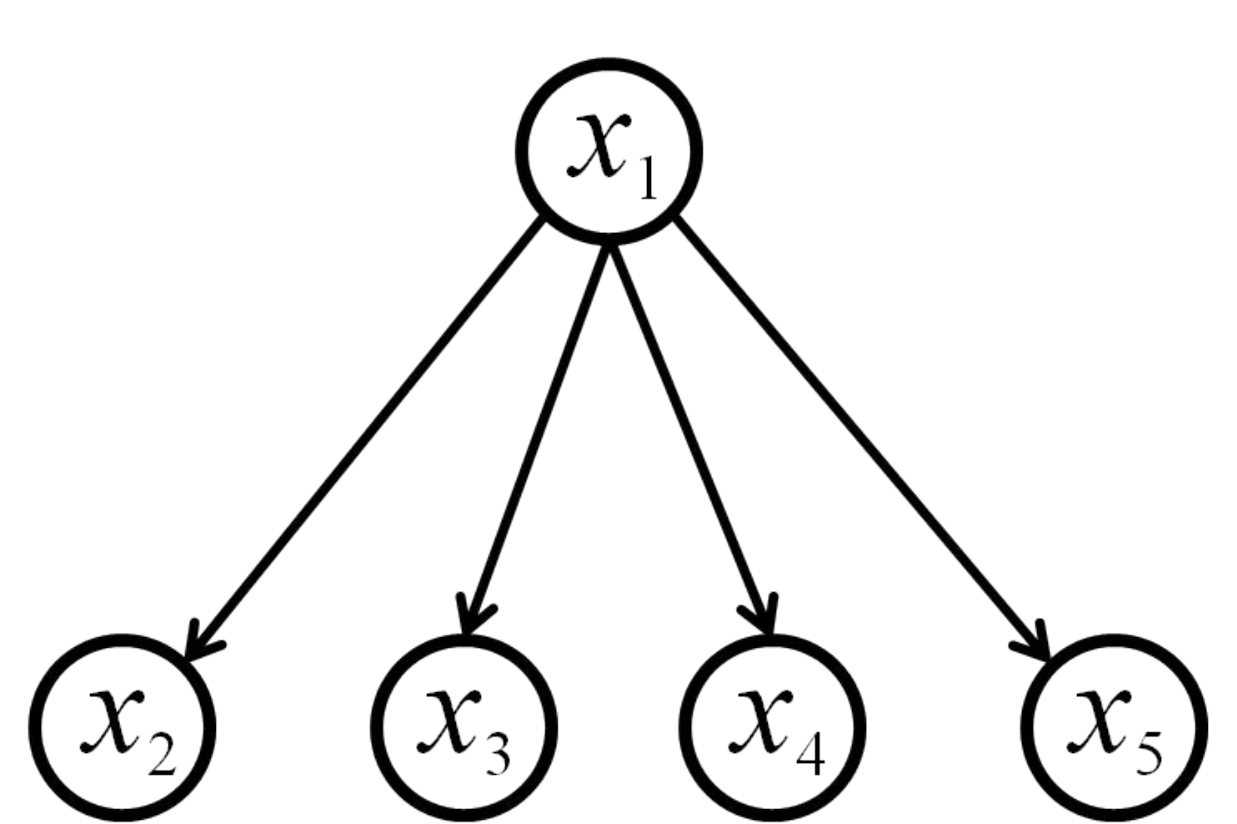}%
\caption{The graph $\cG_3$.}
\label{fig:ex3}%
\end{figure}
The average size of a ball is $\frac{1\cdot 5+4\cdot 1}{5}=9/5$ and thus the average sphere packing value becomes $\frac{5}{9/5} = 25/9$. However, the minimum distance of the code $\cC=\{x_2,x_3,x_4,x_5\}$ in $\cG_3$ is $\infty$, and in particular, it can be a code with minimum distance $3$, which contradicts the average sphere packing value.
\hfill$\Box$
\end{example}

Example~\ref{ex3} depicts a directed, i.e. not symmetric, graph where the average sphere packing value does not hold. Next we show an example of a symmetric graph that does not satisfy the average sphere packing value either.
\begin{example}\label{ex4}
Assume there are $n=k^2$ vertices partitioned into two groups: the first one consists of $k$ vertices and the other group of the remaining $n-k$ vertices. Every vertex from the first group is connected (symmetrically) to a set of exactly $\frac{n-k}{k}=k-1$ vertices from the second group such that there is no overlap between these $k$ sets. The $n-k$ vertices in the second group are all connected to each other. Thus, the average radius-one ball size is
$$\overline{\Delta}_1\hspace{-0.3ex}=\hspace{-0.3ex}\frac{k\cdot k + (n-k)(n-k+1)}{n} \hspace{-0.3ex}=\hspace{-0.3ex} n -2\sqrt{n}+3-\frac{1}{\sqrt{n}} \hspace{-0.3ex}>\hspace{-0.3ex} n/2.$$
Therefore, the average sphere packing value is less than 2. However, it is possible to construct a single-error correcting code with the $k$ vertices of the first group.
\hfill$\Box$
\end{example}

Examples~\ref{ex3} and~\ref{ex4} prove that the average sphere packing value does not hold in all cases. In fact, from Example~\ref{ex4}, we do not only conclude that it does not hold in general, but also that the ratio between this value and a size of a code can be arbitrarily small. However, it is still very interesting to find some minimal conditions such that this bound holds.

\subsection{Monotonicity and Fractional Transversals}\label{subsec:monotonicity}
Remember that a vector $\bfw$ is a fractional transversal if $\bfw\geq \textbf{0}$ and for $1\leq i\leq n$, $$\sum_{y\in B_r(x_i)}w_y\geq 1.$$

A first example for choosing a fractional transversal is stated in the next lemma.
\begin{lemma}\label{lem:first transversal}
The vector $\bfw$ given by
$$w_i = \frac{1}{\min_{x\in B_r^\textmd{in}(x_i)}\{\deg_r(x)\}},$$
for $1\leq i\leq n$, is a fractional transversal.
\end{lemma}
\begin{IEEEproof}
It is easy to verify that $\bfw\geq \textbf{0}$. For every $1\leq i\leq n$, if $y\in B_r(x_i)$, then $x_i\in B_r^\textmd{in}(y)$ and thus
$$w_y = \frac{1}{\min_{x\in B_r^\textmd{in}(y)}\{\deg_r(x)\}} \geq \frac{1}{\deg_r(x_i)}.$$
Therefore, we get
\begin{align*}
& \sum_{y\in B_r(x_i)}w_y \geq \sum_{y\in B_r(x_i)}\frac{1}{\deg_r(x_i)} =1. &
\end{align*}
\end{IEEEproof}

A graph $\cG$ is said to satisfy the \textbf{\emph{monotonicity property}}, or $\cG$ is \textbf{\emph{monotone}}, if for every $r\geq 1$, $x\in X$ and $y\in B_r(x)$,
$$\deg_r(y) \leq \deg_r(x).$$ In this case, the fractional transversal from Lemma~\ref{lem:first transversal} can be stated more explicitly.
\begin{lemma}\label{lem:monotone transversal}
If $\cG$ is monotone then the vector $\bfw$ given by
$$w_i = \frac{1}{\deg_r(x_i)},$$ for $1\leq i\leq n$, is a fractional transversal.
\end{lemma}
\begin{IEEEproof}
If $\cG$ is monotone then for every $x\in B_r^\textmd{in}(x_i)$, $\deg_r(x) \geq \deg_r(x_i)$. Therefore, the fractional transversal $\bfw$ from Lemma~\ref{lem:first transversal} simply becomes
$$
w_i = \frac{1}{\deg_r(x_i)}.
$$
\end{IEEEproof}

As a result of Lemma~\ref{lem:monotone transversal}, if $\cG$ is monotone, then the following expression is an upper bound on $A_\cG(n,2r+1)$,
\begin{equation}\label{eq:monotone upper bound}
A_\cG(n,2r+1)\leq \sum_{i=1}^nw_i = \sum_{i=1}^n\frac{1}{\deg_r(x_i)}.
\end{equation}
We call this bound the \textbf{\emph{monotonicity upper bound}}, which holds in case that $\cG$ is monotone, and denote it by $MB(\cG,r)$. We will build upon Example~\ref{ex:Z1} to exemplify the monotonicity upper bound for the $Z$ channel.
\begin{example}\label{ex:Z2}
It is straightforward to verify that the graph $\cG_Z$ from Example~\ref{ex:Z1} satisfies the monotonicity property since for every $\bfx,\bfy\in\{0,1\}^n$, if $\bfy \in B_{Z,r}$ then $w_H(\bfy)\leq w_H(\bfx)$. Thus, according to Lemma~\ref{lem:monotone transversal}, the vector $\bfw=(w_{\bfx})_{\bfx\in\{0,1\}^n}$ given by
$$w_{\bfx} = \frac{1}{\deg_r(\bfx)} = \frac{1}{\sum_{i=0}^r\binom{w_H(\bfx)}{i}},$$
is a fractional transversal. Therefore, the monotonicity upper bound $MB(\cG_Z,r)$ derived in~(\ref{eq:monotone upper bound}) is calculated to be
\begin{align*}
& MB(\cG_Z,r) = \sum_{\bfx\in\{0,1\}^n}w_{\bfx} = \sum_{\bfx\in\{0,1\}^n}  \frac{1}{\sum_{i=0}^r\binom{w_H(\bfx)}{i}} & \\
& = \sum_{w=0}^n \binom{n}{w}  \frac{1}{\sum_{i=0}^r\binom{w}{i}} &
\end{align*}
For example, for $r=1$, we get
$$ MB(\cG_Z,1)\hspace{-0.3ex} =\hspace{-0.3ex} \sum_{w=0}^n \binom{n}{w}  \frac{1}{\sum_{i=0}^1\binom{w}{i}}\hspace{-0.3ex} =\hspace{-0.3ex} \sum_{w=0}^n \binom{n}{w}  \frac{1}{w+1} = \frac{2^{n+1}}{n+1}.$$
Note that the average sphere packing value, calculated in Example~\ref{ex:Z1}, for $r=1$ is $\frac{2^{n+1}}{n+2}$, is stronger than the monotonicity upper bound. In fact, this hints that in some cases, which will be studied in the sequel, it is possible to improve upon the monotonicity upper bound. Indeed, it is possible to verify that in this case the fractional transversal according to Lemma~\ref{lem:monotone transversal} is not optimal by showing that the vector $\bfw' = (w'_{\bfx})_{\bfx\in\{0,1\}^n}$, where
$$w'_{\bfx} = \frac{1}{w_H(\bfx)+1}\cdot \frac{w_H(\bfx)+2}{w_H(\bfx)+3},$$ for $\bfx\neq \textbf{0}$ and $w'_{\textbf{0}}=1$, is a fractional transversal.
The corresponding bound for this fractional transversal becomes
$$2^{n+1}\cdot \frac{1}{n+3 - \frac{2n+6}{n^2+3n+4}} 
\leq \frac{2^{n+1}}{n+2},$$
which verifies the validity of the average sphere packing value. However, this choice of fractional transversal is still suboptimal and hence we seek to find a further improvement. Finding the exact value $\tau^*(\cH(\cG_Z,r))$ will be the topic and problem we solve in Section~\ref{sec:Z channel}.
\hfill$\Box$
\end{example}

The deletion channel which was studied in~\cite{KK12}, the overlapping grain-error model studied in~\cite{GYD13} and the non-overlapping grain error-error model for $r=1,2,3$ studied in~\cite{KZ13} all satisfy the monotonicity property. Indeed, all these works applied the monotonicity upper bound in order to derive upper bounds on the cardinalities of error-correcting codes in every channel. However, as will be shown in this work, the choice of the fractional transversal according to Lemma~\ref{lem:monotone transversal} is not necessarily optimal. This will be verified by providing different fractional transversals which yield stronger upper bounds than the ones achieved by the monotonicity upper bound.

\subsection{Automorphisms on Graphs}\label{subsec:auto}

One of the main obstacles in calculating the value of $\tau^*(\cH(\cG,r))$ is the large number of variables and constraints in the linear programming in~(\ref{eq:GSPB}). However, most of the graphs studied in this work contain symmetries between their vertices. For example, the linear programming in Example~\ref{ex:Z1} for the $Z$ channel has $2^n$ variables and $2^n$ constraints in order to find the value of $\tau^*(\cH(\cG_Z,r))$, but it is not hard to notice that vectors of the same weight have identical behavior, and thus, one would expect to assign the same weight to these vertices. This will reduce the number of variables and constraints from $2^n$ to $n+1$, which significantly simplifies the linear programming problem in~(\ref{eq:Z GSPB}). This subsection presents a scheme, based upon graph automorphisms, that in many cases can be used in order to significantly reduce the number of variables and constraints to calculate the bound $\tau^*(\cH(\cG,r))$. We will show the general scheme along with a demonstration how it is applied on our continued example of the $Z$ channel.

Let us first remind some tools derived from properties on automorphisms of graphs. Let $G=(X,E)$ be a directed graph with $n$ vertices. An \textbf{\emph{automorphism}} of $G$ is a permutation of its vertices that preserves adjacency. That is, an automorphism of $G$ is a permutation $\pi:X\rightarrow X$ such that for all $(x,y)\in X\times X$, $(x,y)\in E$ if and only if $(\pi(x),\pi(y))\in E$. Assume $|X|=n$, we let $\mathbb{S}_n$ be the set of all permutations of $n$ elements. The set of all automorphisms of $G$ is
$$ Aut(G) = \{ \pi\in \mathbb{S}_n \ |\ \textrm{$\pi$ is an automorphism of $G$}\}.$$ It is known that $Aut(G)$ is a subgroup of the symmetric group $\mathbb{S}_n$ under the operation of functions composition.

The group $Aut(G)$ induces a relation $R$ on $X$ such that $(x,y)\in R$ if and only if there exists $\pi\in Aut(G)$ where $\pi(x)=y$. It is possible to verify that $R$ is an equivalence order and hence $X$ is partitioned into $1\leq n(G)\leq n$ equivalence classes, denoted by $X_1,\ldots,X_{n(\cG)}$. Furthermore, we denote $\cX(G) = \{ X_1,\ldots,X_{n(G)} \}$.

For any $c>0$, let us define the set
$$\cW_c=\bigg\{\bfw \ : \textmd{$\bfw$ is a fractional transversal and }  \sum_{i=1}^nw_{i} = c\bigg\}.$$
Given a partition $\cX =\{X_1,\ldots,X_k\}$ of $X$, we say that a fractional transversal $\bfw$ is \textbf{\emph{$\cX$-regular}} if for all $1\leq j\leq k$ and every $x,y\in X_j$, $w_x = w_y$.

Given a fractional transversal $\bfw$ and an automorphism $\pi \in Aut(G)$, the vector $\bfw^{\pi}$ is defined by $\bfw^{\pi}_i = \bfw_{\pi(i)}$. The next lemma proves that the vector $\bfw^{\pi}$ is a fractional transversal as well.
\begin{lemma}\label{lem:trans}
Let $\bfw$ be a fractional transversal and $\pi$ an automorphism. Then, the vector $\bfw^{\pi}$ is a fractional transversal as well.
\end{lemma}
\begin{IEEEproof}
It is clear to verify that $\bfw^{\pi}\geq \textbf{0}$. We need to show that for all $1\leq i\leq n$ the following inequality holds
$$\sum_{y\in B_r(x_i)}w^{\pi}(y)\geq 1.$$
Since $\pi$ is an automorphism, $y\in B_r(x_i)$ if and only if $\pi(y) \in B_r(\pi(x_i))$ and therefore
$$\sum_{y\in B_r(x_i)}w^{\pi}(y) = \sum_{y\in B_r(x_i)}w_{\pi(y)} = \sum_{y\in B_r(\pi(x_i))}w_{y}\geq 1,$$
where the last inequality holds since $\bfw$ is a fractional transversal.
\end{IEEEproof}

Our main result in this part is stated in the next theorem and corollary.
\begin{theorem}\label{th:regular}
For every $c>0$, if $\cW_c\neq \emptyset$ then $\cW_c$ contains an $\cX(\cG)$-regular fractional transversal.
\end{theorem}
\begin{IEEEproof}
Let $\bfw \in \cW_c $ be a fractional transversal. If $\bfw$ is $\cX(\cG)$-regular then the property holds. Otherwise, let $\pi\in Aut(G)$ and $\bfw^{\pi}$ as defined above. Note that
$$\sum_{i=1}^nw_{i}^{\pi} = \sum_{i=1}^nw_{\pi(i)} = \sum_{i=1}^nw_{i} = c,$$ and together with Lemma~\ref{lem:trans} we get that $\bfw^{\pi}\in \cW_c$. Similarly, we can show that $\frac{\bfw+\bfw^{\pi}}{2}\in  \cW_c$. Let $\pi_1,\pi_2,\ldots,\pi_{N}$ be some order of the automorphisms in $Aut(\cG)$. We can similarly derive that the vector $$\bfw^*=\frac{\sum_{i=1}^{N}\bfw^{\pi_i}}{N}$$ belongs to $\cW_c$ as well.

We finally show that $\bfw^*$ is $\cX(\cG)$-regular. For all $1\leq j\leq n(\cG)$ and $x_{n_1},x_{n_2}\in X_j$
$$w^*_{n_1} = \frac{\sum_{i=1}^Nw^{\pi_i}_{n_1}}{N} = \frac{\sum_{i=1}^Nw_{\pi_i(n_1)}}{N}.$$
Now, let $\pi^*\in Aut(\cG)$ be such that $\pi^*(n_2)=n_1$ and note that
$$\{\pi_1,\ldots,\pi_n\} = \{\pi^*\circ\pi_1,\ldots,\pi^*\circ\pi_n\}.$$
Thus, we get
$$ w^*_{n_2} = \frac{\sum_{i=1}^Nw^{\pi_i}_{n_2}}{N} = \frac{\sum_{i=1}^Nw^{\pi^*\circ\pi_i}_{n_2}}{N} = \frac{\sum_{i=1}^Nw_{(\pi^*\circ\pi_i)(n_2)}}{N}$$
$$ =  \frac{\sum_{i=1}^Nw_{\pi_i(\pi^*(n_2))}}{N} = \frac{\sum_{i=1}^Nw_{\pi_i(n_1)}}{N} = w^*_{n_1}.$$
\end{IEEEproof}

Lastly, we note that Theorem~\ref{th:regular} holds not only for the automorphism group $Aut(\cG)$ but also for every subgroup $H$ of $Aut(\cG)$.
Given a subgroup $H$ of $Aut(\cG)$, assume it partitions the vertices set $X$ into $n_H$ equivalence classes $\cX_H(\cG) =\{X_1,\ldots,X_{n_H}\}$. Let $A_H$ be an $n_H\times n_H$ adjacency matrix corresponding to the subgroup $H$, such that for $1\leq i,j\leq n_H$,
\begin{equation}\label{eq:auth adjaceny mat}
A_H(i,j) = \frac{| \{ (x,y) : x\in X_i, y\in  B_r(x)\cap X_j \} |}{|X_i|}.
\end{equation}
The next Corollary summarizes this discussion.
\begin{corollary}\label{cor:Aut GSPB}
Let $H$ be a subgroup of $Aut(\cG)$ and $\cX_H(\cG)=\{X_1,\ldots,X_{n_H}\}$ is its partition of $X$ into $n_H$ equivalence classes. Then, the generalized sphere packing bound $\tau^*(\cH(\cG,r))$ from~(\ref{eq:GSPB}) becomes
\begin{equation}\label{eq:Aut GSPB}
\tau^*(\cH(\cG,r)) =\min\bigg\{ \sum_{i=1}^{n_H}|X_{i}| w_{i} : A_H^T\cdot \bfw \geq 1, \bfw \in \mathbb{R}^{n_H}_+  \bigg\}.
\end{equation}
\end{corollary}
\begin{IEEEproof}
According to Theorem~\ref{th:regular}, it is enough to consider only fractional transversals which are $\cX_H(\cG)$-regular. Such a fractional transversal can be represented by a vector $\bfw \in \mathbb{R}^{n_H}_+$ such that for $1\leq i\leq n_H$, $w_i$ is the weight given to all the vectors in the set $X_i$.

The condition $A^T\cdot \bfw$ from~(\ref{eq:GSPB}) can be stated as for all $x\in X$, $\sum_{y\in B_r(x)}w_y\geq 1$. However, for all $x\in X_i$ the number of vertices $y\in B_r(x)$ which belong to some set $X_j$ is fixed and is given by the value $A_H(i,j)$. Therefore, for every $x\in X_i$, this condition can be written as $\sum_{j=1}^{n_H} A_H(i,j) w_j \geq 1$. Finally, since there are $|X_i|$ vectors which are assigned with weight $w_i$ we get that the weight of this $\cX_H(\cG)$-regular fractional transversal is $\sum_{i=1}^{n_H}|X_i|w_i$ and thus the corollary holds.
\end{IEEEproof}

The next example shows how to apply the automorphisms scheme presented in this subsection for the $Z$ channel.
\begin{example}\label{ex:Z3}
In Example~\ref{ex:Z1}, we saw that in order to find the value $\tau^*(\cH(\cG_Z,r))$ according to~(\ref{eq:Z GSPB}), it is required to solve a linear programming with $2^n$ variables and $2^n$ constraints. Let us demonstrate how the automorphism scheme studied in this subsection can reduce both the number of variables and constraints to be $n+1$.

First, we define the following set of automorphisms on $\cG_Z$. For every $\sigma\in \mathbb{S}_n$, a permutation $\pi_\sigma:\{0,1\}^n\rightarrow \{0,1\}^n$ is defined such that for all $\bfx\in \{0,1\}^n$, $(\pi_\sigma(\bfx))_i = x_{\sigma(i)}$. It is possible to verify that the set $H=\{\pi_\sigma\ : \ \sigma\in \mathbb{S}_n\}$ is a subgroup of $Aut(\cG_Z)$. Furthermore, the set $\{0,1\}^n$ is partitioned under $H$ into $n+1$ equivalence classes $\cX_H(\cG_Z)=\{X_0,X_1,\ldots,X_n\}$, where $X_i=\{\bfx\in\{0,1\}^n \ : w_H(\bfx)=i \}$, for $0\leq i\leq n$. Therefore, according to equation~(\ref{eq:Aut GSPB}) in Corollary~\ref{cor:Aut GSPB}, it is enough to limit our search and find only fractional transversals $\bfw$ which are $\cX_H(\cG_Z)$-regular.
Hence, the problem in~(\ref{eq:Z GSPB}) is simplified to be

\vspace{-2ex}
\begin{small}
\begin{equation}\label{eq:Z GSPB2}
\hspace{-0.5ex}\tau^*(\cH(\cG_Z,r))\hspace{-0.5ex}=\min\hspace{-0.5ex}\bigg\{\hspace{-0.5ex}\sum_{\ell=0}^n\hspace{-1ex}\binom{n}{\ell}w_{\ell} : \hspace{-1.5ex} \sum_{i=0}^{\min\{\ell,r\}}\hspace{-1ex}\binom{\ell}{i}w_{\ell-i} \geq 1, 0\leq \ell \leq n  \bigg\}.
\end{equation}
\end{small}
\hfill$\Box$
\end{example}

In the next section we will continue Example~\ref{eq:Z GSPB} and show exactly how to solve the problem in~(\ref{eq:Z GSPB2}).

\section{The $Z$ Channel}\label{sec:Z channel}

The $Z$ channel was already discussed before in Examples~\ref{ex:Z1}, \ref{ex:Z2}, and~\ref{ex:Z3}. We derived the linear programming problem to find the value $\tau^*(\cH(\cG_Z,r))$ in~(\ref{eq:Z GSPB}) and calculated its average sphere packing value. Then, we saw that $\cG_Z$ is monotone and thus we calculated its monotonicity upper bound. Finally, we showed how to use the graph automorphism approach in order to derive a more compact linear programming problem to calculate $\tau^*(\cH(\cG_Z,r))$ in~(\ref{eq:Z GSPB2}).

The goal of this section is to solve the linear programming problem in~(\ref{eq:Z GSPB2}) by finding the appropriate fractional transversal and prove that it gives the value of $\tau^*(\cH(\cG_Z,r))$. This result is proved in the next theorem.
\begin{theorem}\label{thm:Z Channel}
For all $r\leq 20$, the optimal fractional transversal which solves the linear programming in ~(\ref{eq:Z GSPB2}) is given by the following recursive formula
\begin{align}
&w_n^*=w_{n-1}^*=\cdots=w_{n-r+1}^*=0,\label{eq:Z recursive}\\
&w_k^*=(1-\sum_{i=1}^{r}w_{k+i}^*\binom{k+r}{r-i})/\binom{k+r}{r}, \forall 1\leq k\leq n-r,\nonumber\\
&w_0^*=1.\nonumber
\end{align}
\end{theorem}
Soon, we will show the equivalent formula
\begin{align}
&w_0^*=1, &\label{eq:Z explicit}\\
&w_k^*=r!k!\sum_{m=r+k}^{n}{D_{m-k-1}\over m!} &\forall k\geq 1,\nonumber
\end{align}
where $D_i$ is given by another recursive relation independent from $n$:
\begin{align}\label{eq:D recursive}
&D_0=D_1=\cdots=D_{r-2}=0,&\nonumber\\
&D_{r-1}=1,&\nonumber\\
&{D_i\over r!}+{D_{i-1}\over (r-1)!}+\cdots+{D_{i-r}\over 0!}=0 &\forall i\geq r.
\end{align}

Furthermore, we note that it is possible to verify the statement for the weight assignment from ~(\ref{eq:Z recursive}) for arbitrary $r$ using the method in theorem \ref{thm:Z Channel}

We divide the proof into three parts. First, we show the equivalence of the two formulas above. Then, we show that $\bf{w}^*$ is in fact a transversal or in other words, it is in the feasibility region of the linear programming. Next, we discuss its optimality. Our method shows both feasibility and optimality for all $r\leq 20$ and we conjecture that $\bf{w}^*$ is the optimal transversal weight for all radius $r\in \mathbb{N}$. One can apply the method to derive the proof for larger $r$.

\subsection{Equivalence of the two formulas}

In order to see the equivalence of two definitions, we fix $r$ and look at $w_k^*$ as a function of both $k$ and $n$ denoted by $w_k^*(n)$ in this subsection. Lets define the sequence $\Delta_k(n)$ as $\Delta_k(n)=w_k^*(n)-w_k^*(n-1)$ for all $n$. So,
\begin{align*}
&\Delta_{k}(n)={1\over \binom{n}{r}} &\text{if } k=n-r,\\
&\Delta_k(n)=0 &\forall k>n-r,\\
&\Delta_k(n)=-\sum_{i=1}^t\Delta_{k+i}(n).{\binom{k+r}{r-i} \over \binom{k+r}{r}} &\forall k< n-r.
\end{align*}

Now, we define another sequence $D_i(n)$ as $D_i(n)=\Delta_{n-i-1}(n).{n!\over r!(n-i-1)!}$ to normalize and reverse the direction of the recursion:
\begin{align*}
&D_0(n)=D_1(n)=\cdots=D_{r-2}(n)=0,\\
&D_{r-1}(n)=1,\\
&{D_i(n)\over r!}+{D_{i-1}(n)\over (r-1)!}+\cdots+{D_{i-r}(n)\over 0!}=0 &\forall i\geq r.
\end{align*}
Note that $D_i(n)$ is independent of $n$. So, we drop $n$ and write $w_k^*(n)$ as
\begin{align*}
&w_k^*(n)=\Delta_{k}(n)+\Delta_{k}(n-1)+\cdots+\Delta_{k}(k+r)\\
&={r!k!\over n!}D_{n-k-1}+{r!k!\over (n-1)!}D_{n-k-2}+\cdots+{r!k!\over (k+r)!}D_{r-1}(n)\\
&=r!k!\sum_{m=r+k}^{n}{D_{m-k-1}\over m!}.
\end{align*}

We can also replace $D_i$ with $\sum_{j=1}^r \alpha_j\lambda_j^i$, where $\lambda_j$'s are roots of the characteristic polynomial $g(x)=\sum_{j=0}^r{x^j\over j!}$ and $\alpha_j$'s are some fixed coefficients found by solving the system of linear equations corresponding to first $r$ initial values.
\subsection{Transversal property for $\bf{w}^*$}

The definition of $\bf{w}^*$ in ~(\ref{eq:Z recursive}) ensures that the inequality constraints in ~(\ref{eq:Z GSPB2}) are satisfied. So, the non-negativity of $\bf{w}^*$ is enough to show $\bf{w}^*$ is a valid transversal.

First, we study the case $r=1$. A simple induction on $i$, shows that $D_i=(-1)^i$. Therefore,
\begin{align*}
w_k^*=&\sum_{m=k+1}^{n}{(-1)^{m-k-1}k!\over m!}\\
=&\bigg({1\over k\hspace{-.4ex}+\hspace{-.4ex}1}-{1\over (k\hspace{-.4ex}+\hspace{-.4ex}1)(k\hspace{-.4ex}+\hspace{-.4ex}2)}\bigg)+\bigg({1\over (k\hspace{-.4ex}+\hspace{-.4ex}1)(k\hspace{-.4ex}+\hspace{-.4ex}2)(k\hspace{-.4ex}+\hspace{-.4ex}3)}\\
&-{1\over (k\hspace{-.4ex}+\hspace{-.4ex}1)(k\hspace{-.4ex}+\hspace{-.4ex}2)(k\hspace{-.4ex}+\hspace{-.4ex}3)(k\hspace{-.4ex}+\hspace{-.4ex}4)}\bigg)\pm\cdots >0.
\end{align*}

In general, it is not easy derive an explicit formula for $\bf{w}^*$ for $r\geq 2$. However, we show that $D_m$ is bounded by an exponential function of $2r$ and hence, the first few terms in ~(\ref{eq:Z explicit}) are dominant comparing to the rest and $w_k^*\geq 0$ is mostly the case. Let us first verify the statement $w_k^*>0$ for $k\geq 3r-1$:
\begin{align}\label{proof:Z 3k-1}
w_k^*&=r!k!\sum_{m=r+k}^{n}{D_{m-k-1}\over m!}\nonumber\\
&=r!k!\bigg({1\over (r\hspace{-.4ex}+\hspace{-.4ex}k)!}+\sum_{m=r+k+1}^{n}{D_{m-k-1}\over m!}\bigg)\nonumber\\
&\geq {r!k!\over (r\hspace{-.4ex}+\hspace{-.4ex}k)!}\bigg(1-\sum_{m=r+k+1}^{n}{|D_{m-k-1}|(r+k)!\over m!}\bigg)\nonumber\\
&\geq {r!k!\over (r\hspace{-.4ex}+\hspace{-.4ex}k)!}\bigg(1-\sum_{m=r+k+1}^{n}{|D_{m-k-1}|\over (4r)^{m-r-k}}\bigg)\nonumber\\
&\geq {r!k!\over (r\hspace{-.4ex}+\hspace{-.4ex}k)!}\bigg(1-\sum_{m=r+k+1}^{n}{(2r)^{m-r-k}\over (4r)^{m-r-k}}\bigg)\\
&= {r!k!\over (r\hspace{-.4ex}+\hspace{-.4ex}k)!}\bigg(1-\sum_{m=1}^{n-k-r}2^{-m}\bigg)={r!k!\over (r\hspace{-.4ex}+\hspace{-.4ex}k)!}2^{-(n-k-r)}>0,\nonumber
\end{align}
where the last inequality comes from Lemma \ref{lemma: bound on D} in appendix \ref{app:Z optimality}. In other words, for $k\geq 3r-1$, the first term in ~(\ref{eq:Z recursive}) is larger than the sum of the absolute values of the remaining terms and they cannot cancel it out. The proof of the case $k<3r-1$ is incomplete for arbitrary radius $r$. However, we introduce a method to verify the feasibility (transversal property) of $\bf{w}^*$ for any fixed $r$ in the following fashion:

Given $k<3r-1$, we look for a number $n_k$ such that
\begin{align}\label{eq:n_k feasibility}
&\sum_{m=r+k}^{n_k}{D_{m-k-1}\over m!}\geq {1\over (2r)^{r+k}}(e^{2r}-\sum_{m=0}^{n_k}{(2r)^m\over m!}),
\end{align}
which means for all $n>n_k$ we have
\begin{align*}
w_k^*&=r!k!\sum_{m=r+k}^{n}{D_{m-k-1}\over m!}\\
&=r!k!(\sum_{m=r+k}^{n_k}{D_{m-k-1}\over m!}+\sum_{m=n_k+1}^{n}{D_{m-k-1}\over m!})\\
&\geq r!k!(\sum_{m=r+k}^{n_k}{D_{m-k-1}\over m!}-\sum_{m=n_k+1}^{n}{(2r)^{m-k-r}\over m!})\\
&> r!k!(\sum_{m=r+k}^{n_k}{D_{m-k-1}\over m!}-{1\over (2r)^{k+r}}\sum_{m=n_k+1}^{\infty}{(2r)^{m}\over m!})\\
&=r!k!(\sum_{m=r+k}^{n_k}{D_{m-k-1}\over m!}-{e^{2r}-\sum_{m=0}^{n_k}{(2r)^m\over m!}\over (2r)^{r+k}})\geq 0;
\end{align*}
And then we check the values of $w_k^*$ for the finite set of $k<3r-1$ and $n\leq n_k$. Note that,
\begin{align*}
\lim_{n_k\rightarrow \infty} e^{2r} - \sum_{m=0}^{n_k}{(2r)^m\over m!}=0.
\end{align*}
Also, $D_i$ is bounded by an exponential function (see Lemma ~\ref{lemma: bound on D}) and hence the following limit exists
\begin{align*}
\ell_k :=\lim_{n_k\rightarrow \infty} \sum_{m=r+k}^{n_k}{D_{m-k-1}\over m!}.
\end{align*}
Finally, if $w_k^*>\epsilon_k> 0$ for all $n>k+r$, then $\ell_k\geq {\epsilon_k\over r!k!}>0$. So, the number $n_k$ should exists. As an example, when $r=2$ we have $n_1=n_2=6, \text{ and } n_3=n_4=7$. Using the above approach, we have verified the feasibility for all $r\leq 20$.

Our calculations also show that $n_k\leq 4r-1$ for all $n\leq 20$. In appendix ~\ref{app:Z optimality}, we prove that $\bf{w}^*$ defined in ~(\ref{eq:Z recursive}), is also the optimal transversal assignment and gives us the best bound using these approach.

In order to evaluate the results, we compared between the different upper bounds for the $Z$ channel. The first bound is the monotonicity bound (MB in short), which was calculated in Example~\ref{ex:Z2}; the second one is the average sphere packing value (ASPV in short), which was calculated in Example~\ref{ex:Z1}; and the third bound is the generalized sphere packing bound (GSPB in short). The best known (to us) upper bound for the $Z$ channel, due to Weber, De Vroedt, and Boekee~\cite{WVB88}, appears in the last column of Table~\ref{table:Z_1}. We see from Table~\ref{table:Z_1} that this bound is better than the GSPB even under optimal weight assignment. However, the bound of~\cite{WVB88} involves solving an integer programming problem, and the authors of~\cite{WVB88} have computed this bound only for $n\leq 23$. In contrast, our bound in Theorem~\ref{thm:Z Channel} is easy to compute for all $n$, and we give its values for $r=1,2,3,4$ up to $n\leq 32$ in Tables~\ref{table:Z_1},~\ref{table:Z_2},~\ref{table:Z_3}, and~\ref{table:Z_4}.
\begin{table}[h]
\begin{center}
\caption{$Z$ channel: upper bounds comparison for $r=1$}\label{table:Z_1}
\begin{tabular}{|c||c|c|c|c|}
  \hline
  $n$ & MB       & ASPV     & GSPB      & \cite{WVB88}  \\ \hline \hline
  5   & 10       & 9        & 8         & 6 \\
  6   & 18       & 16       & 14        & 12 \\
  7   & 32       & 28       & 26        & 18 \\
  8   & 56       & 51       & 47        & 36 \\
  9   & 102      & 93       & 86        & 62 \\
  10  & 186      & 170      & 159       & 117 \\
  11  & 341      & 315      & 295       & 210 \\
  12  & 630      & 585      & 551       & 410 \\
  13  & 1170     & 1092     & 1032      & 786 \\
  14  & 2184     & 2048     & 1940      & 1500 \\
  15  & 4095     & 3855     & 3662      & 2828 \\
  16  & 7710     & 7281     & 6935      & 5430 \\
  17  & 14563    & 13797    & 13170     & 10374 \\
  18  & 27594    & 26214    & 25075     & 19898 \\
  19  & 52428    & 49932    & 47853     & 38008 \\
  20  & 99864    & 95325    & 91514     & 73174 \\
  21  & 190650   & 182361   & 175351    & 140798 \\
  22  & 364722   & 349525   & 336586    & 271953 \\
  23  & 699050   & 671088   & 647131    & 523586 \\
  24  & 1342177  & 1290555   & 1246069   & ? \\
  25  & 2581110  & 2485513   & 2402690   & ? \\
  26  & 4971026  & 4793490   & 4638907   & ? \\
  27  & 9586980  & 9256395   & 8967211   & ? \\
  28  & 18512790 & 17895697  & 17353537  & ? \\
  29  & 35791394 & 34636833  & 33618332  & ? \\
  30  & 69273666 & 67108864  & 65191862  & ? \\
  31  & 134217728& 130150524 & 126535913 & ? \\
  32  & 260301048& 252645135 & 245818070 & ? \\ \hline
\end{tabular}
\end{center}
\end{table}

\begin{table}[h]
\begin{center}
\caption{$Z$ channel: upper bounds comparison for $r=2$}\label{table:Z_2}
\begin{tabular}{|c||c|c|c|c|}
  \hline
  $n$ & MB     & ASPV   & GSPB  & \cite{WVB88}  \\ \hline \hline
  5   & 7      & 5      & 4     & 2 \\
  6   & 12     & 8      & 6     & 4 \\
  7   & 19     & 13     & 9     & 4 \\
  8   & 31     & 21     & 16    & 7 \\
  9   & 51     & 35     & 27    & 12 \\
  10  & 84     & 59     & 46    & 18 \\
  11  & 140    & 101    & 79    & 32 \\
  12  & 238    & 174    & 138   & 63 \\
  13  & 407    & 303    & 243   & 114 \\
  14  & 703    & 532    & 432   & 218 \\
  15  & 1224   & 942    & 772   & 398 \\
  16  & 2151   & 1680   & 1388  & 739 \\
  17  & 3806   & 3013   & 2510  & 1279 \\
  18  & 6780   & 5433   & 4562  & 2380 \\
  19  & 12153  & 9845   & 8327  & 4242 \\
  20  & 21902  & 17924  & 15260 & 8069 \\
  21  & 39672  & 32768  & 28068 & 14374 \\
  22  & 72190  & 60133  & 51802 & 26679 \\
  23  & 131914 & 110740 & 95904 & 50200 \\
  24  & 241977 & 204600 & 178065 & ? \\
  25  & 445447 & 379146 & 331499 & ? \\
  26  & 822696 & 704555 & 618679 & ? \\
  27  & 1524039 & 1312642 & 1157328 & ? \\
  28  & 2831211 & 2451465 & 2169652 & ? \\
  29  & 5273303 & 4588640 & 4075740 & ? \\
  30  & 9845788 & 8607148 & 7670997 & ? \\
  31  & 18424950 & 16176901 & 14463616 & ? \\
  32  & 34553129 & 30460760 & 27317244 & ? \\  \hline
\end{tabular}
\end{center}
\end{table}

\begin{table}[h]
\begin{center}
\caption{$Z$ channel: upper bounds comparison for $r=3$}\label{table:Z_3}
\begin{tabular}{|c||c|c|c|c|}
  \hline
  $n$ & MB    & ASPV   & GSPB  & \cite{WVB88}  \\ \hline \hline
  5   & 7     & 4      & 2     & 2 \\
  6   & 11    & 6      & 3     & 2 \\
  7   & 17    & 9      & 5     & 2 \\
  8   & 26    & 13     & 7     & 4 \\
  9   & 40    & 20     & 11    & 4 \\
  10  & 63    & 31     & 18    & 6 \\
  11  & 99    & 50     & 29    & 8 \\
  12  & 156   & 80     & 48    & 12 \\
  13  & 248   & 130    & 81    & 18 \\
  14  & 400   & 214    & 136   & 34 \\
  15  & 650   & 357    & 231   & 50 \\
  16  & 1066  & 601    & 395   & 90 \\
  17  & 1764  & 1020   & 682   & 168 \\
  18  & 2946  & 1744   & 1186  & 320 \\
  19  & 4960  & 3006   & 2076  & 616 \\
  20  & 8418  & 5216   & 3653  & 1144 \\
  21  & 14395 & 9108   & 6462  & 2134 \\
  22  & 24786 & 15993  & 11486 & 4116 \\
  23  & 42956 & 28232  & 20507 & 7346 \\
  24  & 74902 & 50081   & 36768 & ? \\
  25  & 131345 & 89240  & 66176 & ? \\
  26  & 231537 & 159687 & 119534 & ? \\
  27  & 410164 & 286866 & 216639 & ? \\
  28  & 729924 & 517216 & 393863 & ? \\
  29  & 1304514 & 935722 & 718180 & ? \\
  30  & 2340710 & 1698286 & 1313176 & ? \\
  31  & 4215629 & 3091572 & 2407381 & ? \\
  32  & 7618868 & 5643846 & 4424196 & ? \\  \hline

\end{tabular}
\end{center}
\end{table}

\begin{table}[h]
\begin{center}
\caption{$Z$ channel: upper bounds comparison for $r=4$}\label{table:Z_4}
\begin{tabular}{|c||c|c|c|c|}
  \hline
  $n$ & MB    & ASPV   & GSPB  & \cite{WVB88}  \\ \hline \hline
  5   & 7     & 4      & 2     & 2 \\
  6   & 11    & 5      & 2     & 2 \\
  7   & 17    & 7      & 3     & 2 \\
  8   & 25    & 10     & 4     & 2 \\
  9   & 38    & 15     & 6     & 2 \\
  10  & 58    & 22     & 9     & 4 \\
  11  & 89    & 33     & 14    & 4 \\
  12  & 135   & 49     & 21    & 4 \\
  13  & 207   & 76     & 34    & 6 \\
  14  & 320   & 118    & 54    & 8 \\
  15  & 496   & 185    & 87    & 12 \\
  16  & 774   & 294    & 143   & 16 \\
  17  & 1217  & 472    & 236   & 26 \\
  18  & 1927  & 767    & 393   & 44 \\
  19  & 3073  & 1258   & 660   & 76 \\
  20  & 4939  & 2081   & 1118  & 134 \\
  21  & 7998  & 3470   & 1905  & 229 \\
  22  & 13050 & 5829   & 3266  & 423 \\
  23  & 21450 & 9862   & 5632  & 745 \\
  24  & 35509 & 16791 & 9763 & ? \\
  25  & 59192 & 28761 & 17010 & ? \\
  26  & 99330 & 49540 & 29772 & ? \\
  27  & 167749 & 85775 & 52333 & ? \\
  28  & 285019 & 149239 & 92366 & ? \\
  29  & 487070 & 260846 & 163640 & ? \\
  30  & 836918 & 457873 & 290949 & ? \\
  31  & 1445509 & 806964 & 519048 & ? \\
  32  & 2508896 & 1427610 & 928919 & ? \\  \hline

\end{tabular}
\end{center}
\end{table}

In the next section, we will extend the study of the $Z$ channel for non-binary symbols.

\section{Limited Magnitude Channels}\label{sec:non binary}
We turn in this section to generalize the $Z$ channel for the non-binary case. In this setup, every symbol can have $q$ values, $0,1,\ldots,q-1$ and we denote $[q]=\{0,1,\ldots,q-1\}$. We study the limited magnitude model and focus solely on the single error setup which is carried for two cases. Namely, the error can be asymmetric (Fig.~\ref{fig:non-binary-asym}) or symmetric (Fig.~\ref{fig:non-binary-sym}). This error-channel is motivated by the feature of the errors in non-binary flash memories. The cells in flash memories are charged with electrons and due to the inaccuracy in cell-programming and electrons leakage, the charge level of a cell can either increase or decrease by limited magnitude. For more details see for example~\cite{CSBB10,EB10,KBE10,S14,YMGSSW10}.

\begin{figure}%
\centering
\subfigure[][]{%
\label{fig:non-binary-asym}%
\includegraphics[width =0.5\linewidth]{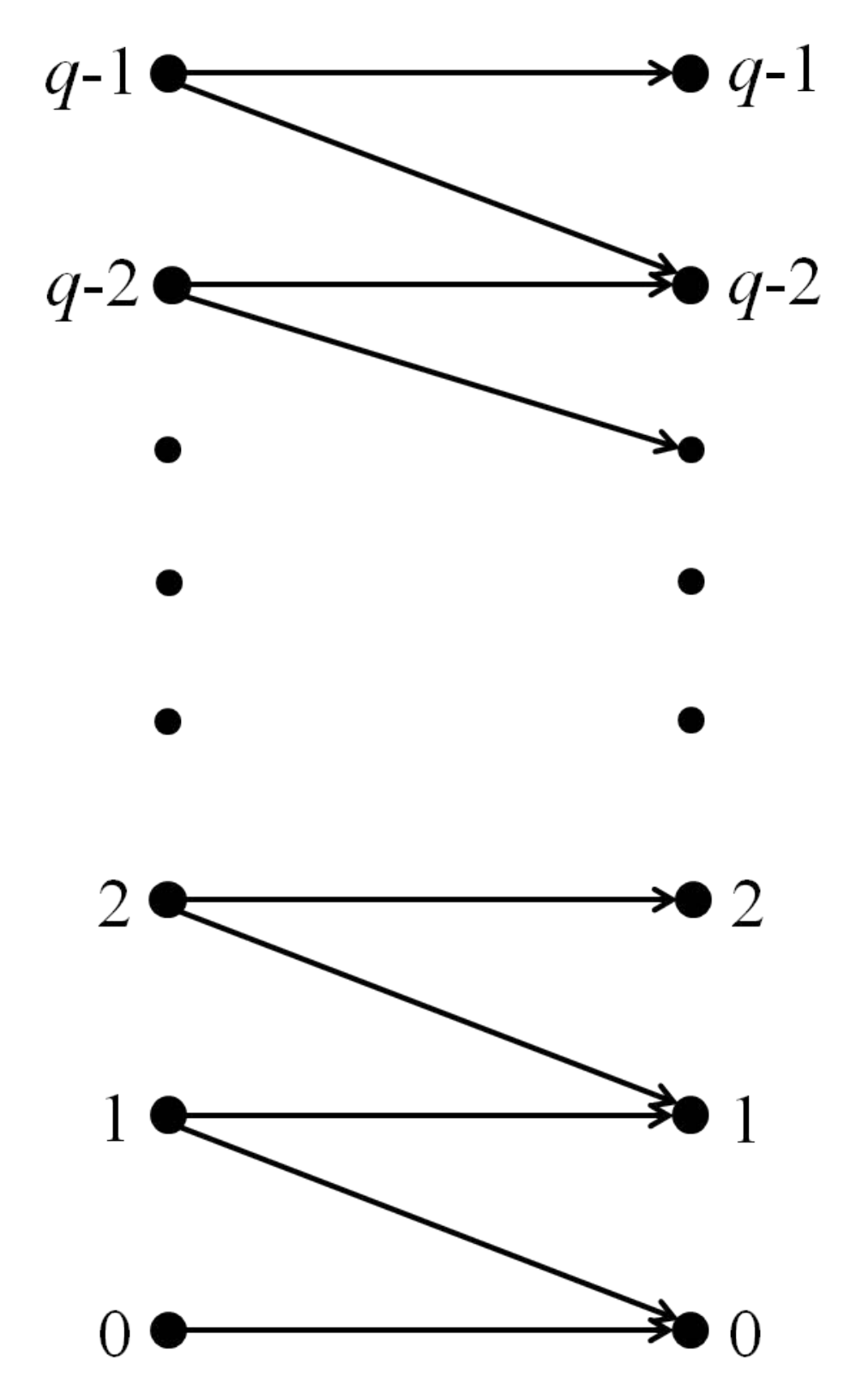}}%
\subfigure[][]{%
\label{fig:non-binary-sym}%
\includegraphics[width =0.5\linewidth]{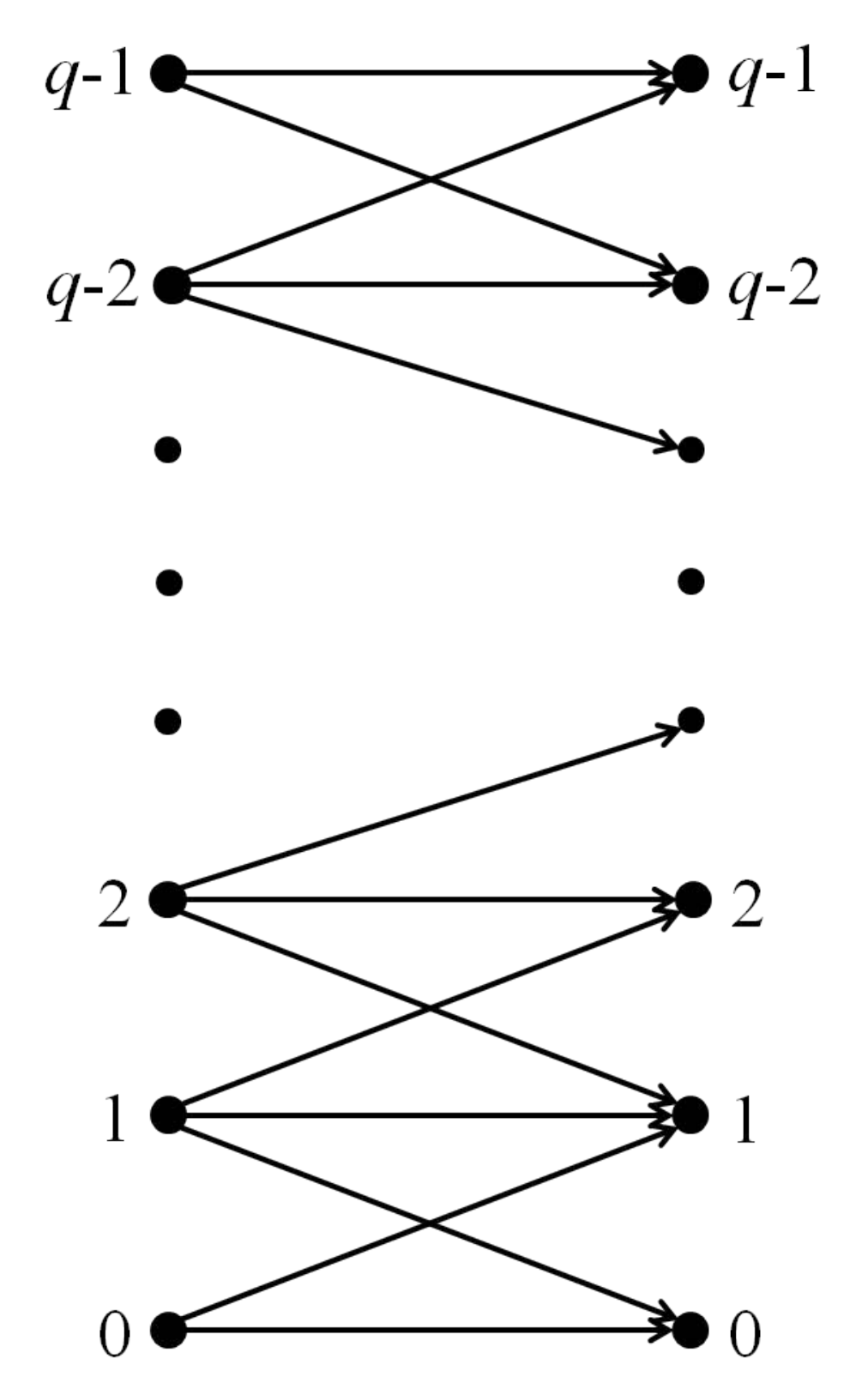}}\\
\caption[Non-Binary Channel.]
{\hspace{-1ex}Two cases of the non-binary channel:
\subref{fig:non-binary-asym} asymmetric errors,
\subref{fig:non-binary-sym} symmetric errors.} 
\label{fig:ex4}%
\end{figure}

\subsection{Asymmetric Errors}\label{subsec:asym}

In the asymmetric non-binary channel, the value of every symbol can only decrease, and in this study we only consider the case where the value of each symbol can decrease by one. The corresponding graph is $\cG_{A,q} = (X_{A,q},E_{A,q})$, where $X_{A,q} = [q]^n$ and
$$E_{A,q} = \bigg\{ (\bfx,\bfy) : \bfx,\bfy\in[q]^n, \bfx \geq \bfy, \sum_{i=1}^nx_i = \sum_{i=1}^ny_i+1 \bigg\}. $$
Given some $\bfx\in [q]^n$, its ball of radius one is described by the set $B_{A,q,1}(\bfx) = \{\bfy\in[q]^n : \bfx\geq \bfy, \sum_{i=1}^nx_i \leq \sum_{i=1}^ny_i+1\}$, and $\deg_{A,q,1}(\bfx) = w_H(\bfx)+1$. The hypergraph in this case is $\cH(\cG_{A,q},1)=(X_{A,q,1},\cE_{A,q,1})$, where $X_{A,q,1} = [q]^n$ and $\cE_{A,q,1} = \{ B_{A,q,1}(\bfx) : \bfx\in[q]^n\}$.

According to the above definitions it is immediate to verify that for all $\bfy\in B_{A,q,1}(\bfx)$, $w_H(\bfy)\leq w_H(\bfx)$ and thus the graph $\cG_{A,q}$ is monotone. In the next two lemmas we calculate the monotonicity upper bound and the average sphere packing value under this setup.
\begin{lemma}\label{lem:MB asym}
The monotonicity upper bound of the graph $\cG_{A,q}$ for $r=1$ is
$$MB(\cG_{A,q},1) = \frac{q^{n+1}}{(q-1)(n+1)}.$$
\end{lemma}
\begin{IEEEproof}
Since the graph $\cG_{A,q}$ is monotone, according to Lemma~\ref{lem:monotone transversal} the following vector $\bfw = (w_{\bfx})_{\bfx\in[q]^n}$ is a fractional transversal,
$$w_{\bfx} = \frac{1}{\deg_{A,q,1}(\bfx)} = \frac{1}{w_H(\bfx)+1}.$$
Thus, the monotonicity upper bound from Equation~(\ref{eq:monotone upper bound}) becomes
\begin{align*}
&  MB(\cG_{A,q},1) =  \sum_{\bfx\in[q]^n}w_{\bfx} = \sum_{\bfx\in[q]^n}\frac{1}{w_H(\bfx)+1} & \\
&  = \sum_{0\leq i_1+\cdots+i_{q-1}\leq n} \binom{n}{i_1,\ldots,i_{q-1}}\frac{1}{i_1+\cdots+i_{q-1}+1} & \\
&  = \frac{q^{n+1}}{(q-1)(n+1)}.&
\end{align*}
\end{IEEEproof}

\begin{lemma}\label{lem:ASPV asym}
The average sphere packing value of the graph $\cG_{A,q}$ for $r=1$ is
$$ASPV(\cG_{A,q},1) = \frac{q^{n+1}}{(q-1)(n+1)+1}.$$
\end{lemma}
\begin{IEEEproof}
The value of the average ball size is
\begin{align*}
&   \frac{1}{q^n}\cdot\sum_{\bfx\in[q]^n}(w_H(\bfx)+1) & \\
& = \frac{1}{q^n}\cdot\sum_{0\leq i_1+\cdots+i_{q-1}\leq n} \binom{n}{i_1,\ldots,i_{q-1}}(i_1+\cdots+i_{q-1}+1) & \\
& = \frac{1}{q^n}\cdot(nq^{n-1} (q-1) + q^n) = n+1 -n/q.&
\end{align*}
Thus, the average sphere packing value in this case is given by
$$\frac{q^n}{n+1 -n/q} = \frac{q^{n+1}}{(q-1)(n+1)+1}.$$
\end{IEEEproof}

The linear programming problem from~(\ref{eq:GSPB}) for this paradigm becomes
$$ \tau^*(\cH(\cG_{A,q,1})) = \min\bigg\{\sum_{\bfx\in[q]^n}w_{\bfx} \ : \ \sum_{\bfy\in B_{A,q,1}(\bfx)}w_{\bfy} \geq 1  \bigg\}.$$
However, it can be significantly simplified according to the tools developed in Section~\ref{subsec:auto}. Similarly to the set of automorphisms from Example~\ref{ex:Z3}, for every permutation $\sigma\in \mathbb{S}_n$ we define a permutation $\pi_{\sigma} = [q]^n\rightarrow [q]^n$ such that for all $\bfx\in[q]^n$, $(\pi_{\sigma}(\bfx))_i = x_{\sigma(i)}$. Hence, also here the set $H_A = \{\pi_\sigma \ : \sigma \in \mathbb{S}_n\}$ is a subgroup of $Aut(\cG_{A,q})$. However, now the subgroup $H_A$ partitions the set $[q]^n$ into the following $n_A =\binom{n+q-1}{q-1}$ equivalence classes
$$\cX = \bigg\{X_{\bfi} \ : \ \bfi = (i_0,\ldots,i_{q-1})\geq \textbf{0}, \sum_{j=0}^{q-1}i_j = n \bigg\},$$
where $X_{\bfi}$ is characterized as follows
$$X_{\bfi} =\{\bfx\in[q]^n : \bfx^{-1}(j) = i_j, 0\leq j\leq q-1 \},$$
and $\bfx^{-1}(j) = |\{ 1\leq k \leq n :x_k = j\}|$.
We denote the set $\bfI_A$ to be $\bfI_A = \{ \bfi \ : \ \bfi = (i_0,\ldots,i_{q-1})\geq \textbf{0}, \sum_{j=0}^{q-1}i_j = n\}$ and define an $n_A\times n_A$ matrix $A_H$ such that its entries are the vectors $(\bfi,\bfj)\in \bfI_A\times \bfI_A$. We assign the values $A_H(\bfi,\bfi) = 1$ and $A_H(\bfi,\bfj) = i_k$ if there exists $1\leq k\leq q-1$ such that $j_k=i_k-1$ and $j_{k-1}=i_{k-1}+1$ and for all $\ell\in[q]\setminus \{k,k-1\}$, $j_{\ell}=i_{\ell}$. All other values in the matrix $A_H$ are assigned with the value 0. Finally, according to Corollary~\ref{cor:Aut GSPB} we proved the following theorem.
\begin{theorem}\label{th:asym frac trans}
The generalized sphere packing bound for $\cH(\cG_{A,q,1})$ is given by
$$\hspace{-1.2ex}\tau^*\hspace{-0.3ex}(\cH(\cG_{A,q,1}))\hspace{-0.5ex} =\hspace{-0.5ex}\min\hspace{-0.5ex}\bigg\{\hspace{-0.5ex}\sum_{\bfi \in \bfI_A}\hspace{-1ex}|X_{\bfi}|w_{\bfi}\hspace{-0.3ex} :\hspace{-0.5ex}A_H^T\hspace{-0.3ex}\cdot\hspace{-0.3ex} \bfw \hspace{-0.5ex}\geq\hspace{-0.5ex} 1, \bfw\hspace{-0.5ex}=\hspace{-0.5ex}(w_{\bfi})_{\bfi\in \bfI_A}\hspace{-1ex} \in\hspace{-0.5ex} \mathbb{R}^{n_A}_+ \hspace{-0.5ex}\bigg\}.$$
\end{theorem}

We finish this section by showing an improvement upon the suboptimal monotonicity upper bound from Lemma~\ref{lem:MB asym}.
In the fractional transversal notation of Theorem~\ref{th:asym frac trans}, if one applied the monotonicity upper bound, then the fractional transversal assignment would be $w_{\bfi} = 1/(n-i_0+1)$ for $\bfi\in \bfI_A$. However, under this assignment almost all of the constraints hold with strict inequality. We show that it is possible to reduce the weights in this assignment without violating the constraints and thus receive a stronger upper bound.
\begin{theorem}\label{th:asym upper improve}
The vector $\bfw = (w_{\bfi})_{\bfi\in \bfI_A}$ given by
$$ w_{\bfi} = \frac{1}{n-i_0+1+\frac{i_1-1}{2(n-i_0)}},$$ if $i_0\neq n$ and otherwise $ w_{\bfi} =1$
is a fractional transversal for $\tau^*(\cH(\cG_{A,q,1}))$ as stated in Theorem~\ref{th:asym frac trans}.
\end{theorem}
\begin{IEEEproof}
It is straightforward to verify that $w_{\bfi}\geq 0$ for all $\bfi\in \bfI_A$. According to the conditions for $\tau^*(\cH(\cG_{A,q,1}))$ from Theorem~\ref{th:asym frac trans}, we need to show that for all $\bfi = (i_0,i_1,\ldots,i_{q-1}) \in \bfI_A$ the following inequality holds
\begin{align*}
& w_{(i_0,i_1,\ldots,i_{q-1})} + i_1w_{(i_0+1,i_1-1,\ldots,i_{q-1})} + i_2w_{(i_0,i_1+1,i_2-1\ldots,i_{q-1})} & \\
& + \cdots + i_{q-1}w_{(i_0,i_1,\ldots,i_{q-2}+1,i_{q-1}-1)}\geq 1. &
\end{align*}
If $i_0=n$ then this inequality holds with equality and it is possible to verify that it holds for $i_0=n-1$ as well. Thus, we can assume that $i_0<n-1$. After placing the values of $w_{\bfi}$ stated in the theorem, we need to show the following
\begin{align*}
& \frac{1}{n-i_0+1+\frac{i_1-1}{2(n-i_0)}} + \frac{i_1}{n-i_0+\frac{i_1-2}{2(n-i_0-1)}} & \\
& + \frac{i_2}{n-i_0+1+\frac{i_1}{2(n-i_0)}} + \frac{i_3+\cdots +i_{q-1}}{n-i_0+1+\frac{i_1-1}{2(n-i_0)}}\geq 1.&
\end{align*}
Note that
$$\frac{1}{n-i_0+1+\frac{i_1-1}{2(n-i_0)}} \geq \frac{1}{n-i_0+1+\frac{i_1}{2(n-i_0)}},$$
and thus it is enough to show that
\begin{align*}
& \frac{i_2+i_3+\cdots +i_{q-1}+1}{n-i_0+1+\frac{i_1}{2(n-i_0)}} + \frac{i_1}{n-i_0+\frac{i_1-2}{2(n-i_0-1)}} \geq 1&
\end{align*}
or, since $i_0+i_1+\cdots + i_{q-1}=n$,
\begin{align*}
& \frac{n-i_0-i_1+1}{n-i_0+1+\frac{i_1}{2(n-i_0)}} + \frac{i_1}{n-i_0+\frac{i_1-2}{2(n-i_0-1)}} \geq 1&
\end{align*}
and
\begin{align*}
& \frac{i_1}{n-i_0+\frac{i_1-2}{2(n-i_0-1)}} \geq 1 - \frac{n-i_0-i_1+1}{n-i_0+1+\frac{i_1}{2(n-i_0)}} &
\end{align*}
which is
\begin{align*}
& \frac{i_1}{n-i_0+\frac{i_1-2}{2(n-i_0-1)}} \geq  \frac{i_1+\frac{i_1}{2(n-i_0)}}{n-i_0+1+\frac{i_1}{2(n-i_0)}}.&
\end{align*}
Let us denote $n-i_0=M$ and we need to show that
\begin{align*}
& \frac{1}{M+\frac{i_1-2}{2(M-1)}} \geq  \frac{1+\frac{1}{2M}}{M+1+\frac{i_1}{2M}},&
\end{align*}
and equivalently
\begin{align*}
& \frac{2(M-1)}{2M^2-2M+i_1-2} \geq \frac{2M+1}{2M^2+2M+i_1},&
\end{align*}
or
\begin{align*}
& 2M^2+2M+2 \geq 3i_1,&
\end{align*}
which holds since $M=n-i_0\geq i_1$.
\end{IEEEproof}

Table~\ref{table:asym_q_3} summarizes the upper bounds results we derived in this section for $q=3$. The first column is the monotonicity upper bound we found in Lemma~\ref{lem:MB asym}. The second column is the average sphere packing value from Lemma~\ref{lem:ASPV asym}. The third column is the improvement in Theorem~\ref{th:asym upper improve} over the monotonicity upper bound. Lastly, the last column is the value of the generalized sphere packing bound from Theorem~\ref{th:asym frac trans}, which we solved numerically. Note that there is no upper bound we know of in the literature for this error channel.
\begin{table}[h]
\begin{center}
\caption{Non-binary channel, asymmetric errors: upper bounds comparison for $q=3$}\label{table:asym_q_3}
\begin{tabular}{|c||c|c|c|c|}
  \hline
  $n$ & MB          & ASPV       & Theorem~\ref{th:asym upper improve}  & GSPB         \\ \hline \hline
  5   & 60          & 56         &  60          & 55           \\
  6   & 156         & 145        &  154         & 144          \\
  7   & 410         & 385        &  402         & 381          \\
  8   & 1093        & 1035       &  1071        & 1021         \\
  9   & 2952        & 2811       &  2888        & 2770         \\
  10  & 8052        & 7702       &  7877        & 7591         \\
  11  & 22143       & 21252      &  21673       & 20955        \\
  12  & 61320       & 59049      &  60056       & 58235        \\
  13  & 170820      & 164929     &  167424      & 162744       \\
  14  & 478296      & 462867     &  469156      & 456987       \\
  15  & 1345210     & 1304446    &  1320524     & 1288583      \\
  16  & 3798240     & 3689718    &  3731321     & 3646657      \\
  17  & 10761680    & 10470824   &  10579575    & 10353898     \\
  18  & 30585828    & 29801576   &  30088394    & 28464819     \\
  19  & 87169610    & 85043521   &  85805885    & 84168158     \\
  20  & 249056028   & 243264027  &  245304388   & 230986164    \\
  21  & 713205900   & 697356880  &  702851238   & 690706260    \\
  22  & 2046590844  & 2003046358 &  2017923470  & 1984633746   \\
  23  & 5883948676  & 5763868091 &  5804351676  & 5712720517   \\ \hline
\end{tabular}
\end{center}
\end{table}

\subsection{Symmetric Errors}\label{subsec:sym}
Since this model and graph are very similar to the asymmetric case, they are briefly presented. The graph is given by $\cG_{S,q} =(X_{S,q},E_{S,q})$, where $X_{S,q} = [q]^n$ and $$E_{S,q} = \{(\bfx,\bfy) : (\bfx,\bfy)\in E_{A,q} \textmd{ or } (\bfy,\bfx)\in E_{A,q}\}.$$
Similarly, for every $\bfx\in[q]^n$, its corresponding ball of radius one is the set
$B_{S,q,1}(\bfx) = \{\bfy\in[q]^n :  \bfy\in  B_{A,q,1}(\bfx) \textmd{ or } \bfx\in  B_{A,q,1}(\bfy)\}$. The hypergraph is $\cH(\cG_{S,q},1)=(X_{S,q,1},\cE_{S,q,1})$, where $X_{S,q,1} = [q]^n$ and $\cE_{S,q,1} = \{ B_{S,q,1}(\bfx) : \bfx\in[q]^n\}$.

This setup is different than all other error channels studied so far in the sense that it does not satisfy the monotonicity property. Thus, we cannot conclude the corresponding fractional transversal of the monotonicity upper bound. However, we can still calculate the average sphere packing value.
\begin{lemma}\label{lem:ASPV_sym}
The average sphere packing value of the graph $\cG_{S,q}$ for $r=1$ is
$$ASPV(\cG_{S,q},1) = \frac{q^n}{2n+1 -2n/q}.$$
\end{lemma}
\begin{IEEEproof}
First we calculate the value of the expected ball size, which is given by
\begin{align*}
&   \frac{1}{q^n}\cdot\sum_{\bfx\in[q]^n}(2n+1-\bfx^{-1}(i_0)-\bfx^{-1}(i_{q-1})) & \\
& = \frac{1}{q^n}\cdot\hspace{-1.5ex}\sum_{0\leq i_1+\cdots+i_{q-1}\leq n}\hspace{-1ex} \binom{n}{i_1,\ldots,i_{q-1}}(n+1+(i_1+\cdots+i_{q-2})) & \\
& = \frac{1}{q^n}\cdot((n+1)q^{n} +n(q-2)q^{n-1}) = 2n+1 -2n/q.&
\end{align*}
Thus, the average sphere packing value becomes
$$\frac{q^n}{2n+1 -2n/q}.$$
\end{IEEEproof}

Next, we define the set of automorphisms to be used here. One can verify that every permutation in $H_A$ is an automorphism in $\cG_{S,q}$. However, in this case we can expand and use more automorphisms. For every binary vector $\bfb\in\{0,1\}^n$, we define the permutation
$$\pi_{\bfb}:[q]^n \rightarrow [q]^n,$$
as follows. For every $\bfx\in [q]^n$, $\pi_{\bfb}(\bfx)$ is the vector defined as
\begin{displaymath}
\pi_{\bfb}(\bfx)_i = \left\{ \begin{array}{ll}
q-1-x_i & \textrm{if $b_i =1$}\\
x_i & \textrm{if $b_i =0$}
\end{array} \right.
\end{displaymath}
Then, the set $H_S = H_A \cup \{\pi_{\bfb} \ : \ \bfb\in\{0,1\}^n\}$ is a subgroup of $Aut(\cG_{S,q})$. The subgroup $H_S$ partitions the set $[q]^n$ into $n_S=\binom{n+\lceil q/2\rceil-1}{\lceil q/2\rceil-1}$ equivalence classes
$$\cX = \bigg\{X_{\bfi} \ : \ \bfi = (i_0,\ldots,i_{\lceil q/2\rceil-1})\geq \textbf{0}, \sum_{j=0}^{\lceil q/2\rceil-1}i_j = n \bigg\},$$
and $X_{\bfi}$ is the set
$$\hspace{-0.5ex}X_{\bfi}\hspace{-0.3ex}=\hspace{-0.3ex}\{\bfx\in[q]^n\hspace{-0.5ex}: \hspace{-0.3ex}\bfx^{-1}\hspace{-0.4ex}(j) \hspace{-0.3ex}+\hspace{-0.3ex} \bfx^{-1}\hspace{-0.4ex}(q-1-j)\hspace{-0.5ex}= i_j, 0\leq\hspace{-0.5ex}j\hspace{-0.5ex}\leq\hspace{-0.5ex} \lceil q/2\rceil-1 \}.$$
We define $$\bfI_S = \{\bfi \ : \ \bfi=(i_0,\ldots,i_{\lceil q/2\rceil-1})\geq \textbf{0}, \sum_{j=0}^{\lceil q/2\rceil-1}i_j = n\}$$
and the $n_S\times n_S$ matrix $A_S$ with the following entries $(\bfi,\bfj)\in \bfI_S\times \bfI_S$.
\begin{enumerate}
\item For all $\bfi\in \bfI_S$, $A_S(\bfi,\bfi)=1$,
\item $A_S(\bfi,\bfj)=k$ if there exists $1\leq k\leq \lceil q/2\rceil-1$ such that $j_k=i_k-1$ and $j_{k-1}= i_{k-1}+1$ and for all $\ell\in [\lceil q/2\rceil]\setminus \{k,k-1\}$, $j_{\ell}=i_{\ell}$.
\end{enumerate}
To conclude, according to Corollary~\ref{cor:Aut GSPB}, the generalized sphere packing for $\cH(\cG_{S,q,1})$ becomes
$$\hspace{-1.2ex}\tau^*\hspace{-0.3ex}(\cH(\cG_{S,q,1}))\hspace{-0.5ex} =\hspace{-0.5ex}\min\hspace{-0.5ex}\bigg\{\hspace{-0.5ex}\sum_{\bfi \in \bfI_S}\hspace{-1ex}|X_{\bfi}|w_{\bfi}\hspace{-0.3ex} :\hspace{-0.5ex}A_S^T\hspace{-0.3ex}\cdot\hspace{-0.3ex} \bfw \hspace{-0.5ex}\geq\hspace{-0.5ex} 1, \bfw\hspace{-0.5ex}=\hspace{-0.5ex}(w_{\bfi})_{\bfi\in \bfI_S}\hspace{-1ex} \in\hspace{-0.5ex} \mathbb{R}^{n_S}_+ \hspace{-0.5ex}\bigg\}.$$

Even though this graph does not satisfy the monotonicity property we can still derive a similar bound, which will be stated in the following theorem.
\begin{theorem}\label{thm:symm upper bound}
The vector $\bfw = (w_{\bfx})_{\bfx\in[q]^n}$ given by $$w_{\bfx} = \frac{1}{\deg_{S,q,1}(\bfx)-1}$$
is a fractional transversal.
\end{theorem}
\begin{IEEEproof}
Let $\bfx\in[q]^n$ and let $i_j = \bfx^{-1}(j)$ for $j\in [q]$. Then, $\deg_{S,q,1}(\bfx) = 2n-i_0-i_{q-1}+1$. We need to show that
\begin{align*}
& \frac{1}{2n-i_0-i_{q-1}} + \frac{i_0}{2n-(i_0-1)-i_{q-1}} & \\
+ & \frac{i_1}{2n-(i_0+1)-i_{q-1}} + \frac{i_1+2i_2+\cdots+ 2i_{q-3}+i_{q-2}}{2n-i_0-i_{q-1}} & \\
+ &  \frac{i_{q-2}}{2n-i_0-(i_{q-1}+1)} + \frac{i_{q-1}}{2n-i_0-(i_{q-1}-1)} \geq 1,&
\end{align*}
or
\begin{align*}
& \frac{1+i_1+2i_2+\cdots+ 2i_{q-3}+i_{q-2}}{2n-i_0-i_{q-1}} + \frac{i_0+i_{q-1}}{2n-i_0-i_{q-1}+1} & \\
+& \frac{i_1+i_{q-2}}{2n-i_0-i_{q-1}-1} \geq 1.&
\end{align*}
Since $1/(2n-i_0-i_{q-1}) \geq 1/(2n-i_0-i_{q-1}+1)$ and $1/(2n-i_0-i_{q-1}-1) \geq 1/(2n-i_0-i_{q-1}+1)$, it is enough to show that
\begin{align*}
& \frac{1+i_1+2i_2+\cdots+ 2i_{q-3}+i_{q-2}}{2n-i_0-i_{q-1}+1} + \frac{i_0+i_{q-1}}{2n-i_0-i_{q-1}+1} & \\
+& \frac{i_1+i_{q-2}}{2n-i_0-i_{q-1}+1} \geq 1,&
\end{align*}
which holds with equality.
\end{IEEEproof}

Comparison results for $q=3$ and $q=4$ are summarized in Tables~\ref{table:sym_q_3} and~\ref{table:sym_q_4}. The first column is the average sphere packing value which was calculated in Lemma~\ref{lem:ASPV_sym}. The second column is the upper bound we found in Theorem~\ref{thm:symm upper bound}. The last column is the value of the generalized sphere packing bound that we solved numerically. Note that in this example the value of the average sphere packing value is less than the one of the generalized sphere packing value, however, that doesn't mean that it is not a valid upper bound.
\begin{table}[h]
\begin{center}
\caption{Non-binary channel, symmetric errors: upper bounds comparison for $q=3$}\label{table:sym_q_3}
\begin{tabular}{|c||c|c|c|}
  \hline
  $n$ & ASPV       & Theorem~\ref{thm:symm upper bound}  & GSPB    \\ \hline \hline
  5   & 31         &  37        &  32       \\
  6   & 81         &  93        &  82       \\
  7   & 211        &  238       &  216      \\
  8   & 562        &  624       &  572      \\
  9   & 1514       &  1663      &  1538     \\
  10  & 4119       &  4484      &  4177     \\
  11  & 11307      &  12217     &  11449    \\
  12  & 31261      &  33564     &  31618    \\
  13  & 86963      &  92872     &  87872    \\
  14  & 243201     &  258535    &  245544   \\
  15  & 683281     &  723466    &  689388   \\
  16  & 1927465    &  2033685   &  1943532  \\
  17  & 5456626    &  5739520   &  5499244  \\
  18  & 15496819   &  16255303   &  15610684 \\ \hline
\end{tabular}
\end{center}
\end{table}

\begin{table}[h]
\begin{center}
\caption{Non-binary channel, symmetric errors: upper bounds comparison for $q=4$}\label{table:sym_q_4}
\begin{tabular}{|c||c|c|c|}
  \hline
  $n$ & ASPV        & Theorem~\ref{thm:symm upper bound}  & GSPB         \\ \hline \hline
  5   & 120         & 139   & 123           \\
  6   & 409         & 463   & 417          \\
  7   & 1424        & 1586   & 1449          \\
  8   & 5041        & 5540   & 5115         \\
  9   & 18078       & 19666   & 18313         \\
  10  & 65536       & 70707   & 66297         \\
  11  & 239674      & 256844   & 242193        \\
  12  & 883011      & 940934   & 891482        \\
  \hline
\end{tabular}
\end{center}
\end{table}

\section{Deletion and Grain-Error Channels}\label{sec:deletions n grains}
In this section we shift our attention to the deletion channel, which was the original usage of the generalized sphere packing bound in~\cite{KK12}. We will only focus on the single-deletion case. First, we revisit the fractional transversal given in~\cite{KK12} to verify that the graph in the deletion channel satisfies a similar property to the monotonicity property from Section~\ref{subsec:monotonicity}. However, it will be noticed that this choice of fractional transversal is suboptimal and thereof an improvement will be presented. This will be our main result in this section, namely, an explicit expression of a fractional transversal which improves upon the one from~\cite{KK12}. Since the structure of the deletion and grain-error channels is very similar, especially for a single error, in the second part of this section we show also how to improve upon the upper bound from~\cite{GYD13,KZ13} on the cardinality of single grain-error-correcting codes.

\subsection{Deletions}
As studied in the previous examples and sections, we first introduce the graph for the deletion channel. However, note that the graph in this setup is different than the previous ones studied so far. Specifically, a length $n$ vector which suffers a single deletion will result with a vector of length $n-1$. To accommodate this structure, the vertices in the graph are defined to be both vectors of length $n$ and $n-1$, so the graph is $\cG_{D} = (X_{D},E_{D})$, where $X_{D} = \{0,1\}^n\cup \{0,1\}^{n-1}$ and
\begin{align*}
E_D = \{ & (\bfx,\bfy)\in \{0,1\}^n\times \{0,1\}^{n-1} : & \\
         & \bfy = (x_1,\ldots,x_i,x_{i+2},\ldots,x_n)\textmd{ for some }1\leq i\leq n  \}. &
\end{align*}
For any $\bfx\in \{0,1\}^n$, its radius one ball is the set $B_{D,1}(\bfx) = \{\bfy\in\{0,1\}^{n-1} : (\bfx,\bfy)\in E_D\}$, and for $\bfx\in \{0,1\}^{n-1}$, $B_{D,1}(\bfx) = \emptyset$. Therefore, $1\leq \deg_{D,1}(\bfx) \leq n$ for $\bfx\in \{0,1\}^n$, and $\deg_{D,1}(\bfx) = 0$ for $\bfx\in \{0,1\}^{n-1}$.

At this point, we could basically construct the hypergraph for the deletion channel as was done in the previous examples such that its set of vertices is $X_{D} = \{0,1\}^n\cup \{0,1\}^{n-1}$. However, since the length-$n$ vectors do not participate in the balls we can eliminate them in the hypergraph construction, which coincides with the hypergraph construction in~\cite{KK12}. Thus the hypergraph for the single deletion channel is $\cH(\cG_{D},1)=(X_{D,1},\cE_{D,1})$, where $X_{D,1} = \{0,1\}^{n-1}$ and $\cE_{D,1} = \{B_{D,1}(\bfx) : \bfx\in \{0,1\}^{n}\}$. This definition does not change the analysis of the upper bounds studied in this paper.
Thus, the generalized sphere packing bound in this setup becomes
\begin{equation}\label{eq:GSPB deletion}
\hspace{-1ex}\tau^*(\cH(\cG_{D},1))\hspace{-0.5ex}=\hspace{-0.5ex}\min\hspace{-0.5ex} \bigg\{\hspace{-0.6ex}\sum_{\bfz\in\{0,1\}^{n-1}}\hspace{-3ex}w_{\bfz}  : \hspace{-1.2ex} \sum_{\bfy\in B_{D,1}(\bfx)}\hspace{-2.5ex}w_{\bfy} \geq 1, \forall \bfx\in\{0,1\}^n \bigg\}.
\end{equation}

For a vector $\bfx\in\{0,1\}^n$, we denote by $\rho(\bfx)$ the number of runs in $\bfx$. For example, if $\bfx = 001010010$, then $\rho(\bfx)=7$. It is easily verified that for $\bfx\in\{0,1\}^n$, $\deg_{D,1}(\bfx) = \rho(\bfx)$,~\cite{KK12}. It is also known that the number of length-$n$ vectors with $1\leq \rho \leq n$ runs is $2\binom{n-1}{\rho-1}$.
Let us first calculate the average sphere packing value for the hypergraph $\cH(\cG_{D},1)$. This will be done in the next lemma.
\begin{lemma}\label{lem:D ASPV}
The average sphere packing value of the graph $\cG_{D}$ for $r=1$ is
$$ASPV(\cG_{D},1) = \frac{2^{n}}{n+1}.$$
\end{lemma}
\begin{IEEEproof}
Every vector $\bfx\in\{0,1\}^n$ generates a ball, i.e. a hyperedge, in $\cH(\cG_{D},1)$. Thus, the average size of a ball is given by
\begin{align*}
& \frac{1}{2^n}\sum_{\bfx\in\{0,1\}^n} \deg_{D,1}(\bfx) = \frac{1}{2^n}\sum_{\rho=1}^n 2\binom{n-1}{\rho-1} \rho & \\
& = \frac{1}{2^{n-1}}\sum_{\rho=0}^{n-1} \binom{n-1}{\rho} (\rho+1) = \frac{1}{2^{n-1}}(2^{n-1}+(n-1)2^{n-2})& \\
& = \frac{n+1}{2}. &
\end{align*}
Thus, the average sphere packing value becomes
$$\frac{2^{n-1}}{(n+1)/2} = \frac{2^{n}}{n+1}.$$
\end{IEEEproof}
Note that if one chose the hypergraph to contain all binary vectors of length $n-1$ and $n$, the resulting average sphere packing value would have been weaker. We specifically chose the hypergraph this way as it is the smallest one where any single-deletion code can be studied and analyzed.

In the setup and structure of the graph $\cG_D$, it is not possible to indicate whether the graph $\cG_D$ satisfies the monotonicity property. The vectors in the ball centered at some $\bfx\in\{0,1\}^n$ are of length $n-1$ and thus do not have corresponding balls. However, there is still a very similar property to the monotonicity one. Namely, for every $\bfy\in B_{D,1}(\bfx)$, where $\bfx\in \{0,1\}^n$,
$$\rho (\bfy) \leq \rho(\bfx) = \deg_{D,1}(\bfx).$$
This property was established in~\cite{KK12} and thus a choice of a fractional transversal $(w_{\bfx})_{\bfx\in \{0,1\}^{n-1}}$, was given by
$$w_{\bfx}=\frac{1}{\rho(\bfx)}.$$
The corresponding upper bound, which we call here the monotonicity upper bound, was calculated in~\cite{KK12} to be
$$\sum_{\bfx\in\{0,1\}^{n-1}} \frac{1}{\rho(\bfx)} = \sum_{\rho=1}^{n-1}2\binom{n-2}{\rho-1}\cdot \frac{1}{\rho}= \frac{2^n-2}{n-1}.$$

However, it is possible to verify that for this fractional transversal many of the constraints in the linear programming in~(\ref{eq:GSPB deletion}) hold with strong inequality, which implies that a better one could be found. This will be the focus in the rest of this subsection, that is, an improvement upon the last upper bound by the equivalent of the monotonicity property.

For a vector $\bfx$, let $\mu(\bfx)$ be the number of middle runs (i.e., not on the edges) of length 1 in $\bfx$. We call these runs \emph{middle-1-runs}. For example, for $\bfx = 001010010$, $\mu(\bfx) = 4$. First notice that if $\rho(\bfx)\geq 2$ then $0\leq \mu(\bfx)\leq \rho(\bfx)-2$. Let $N_n(\rho,\mu)$ denote the number of vectors of length $n$ with $\rho$ runs and $\mu$ middle-1-runs. For $\rho=1$ and $\mu=0$, we have $N_n(1,0)=2$. For $2\leq \rho\leq n$ and $0\leq \mu \leq \rho-2$, the value of $N_n(\rho,\mu)$ is calculated in the next lemma. For all other values of $\rho$ and $\mu$ the value of $N_n(\rho,\mu)$ is zero.
\begin{lemma}\label{lem:count}
For $2\leq \rho\leq n$ and $0\leq \mu \leq \rho-2$,
$$N_n(\rho,\mu)=2\binom{\rho-2}{\mu}\binom{n-\rho+1}{\rho-\mu-1}.$$
\end{lemma}
\begin{IEEEproof}
For every $\bfx=(x_1,\ldots,x_n)\in\{0,1\}^n$, let $\bfx'=(x_1',\ldots,x_{n-1}')\in\{0,1\}^{n-1}$ be a vector of length $n-1$ such that for $1\leq i\leq n-1$, $x_i'=x_i+x_{i+1}$. Note that $w_H(\bfx') = \rho(\bfx)-1$. Let $c(\bfx')$ denote the number of times that two consecutive ones appear in $\bfx'$, so we have $c(\bfx') = \mu(\bfx)$.

Let $\bfx$ be a vector of length $n$ such that $\rho(\bfx) = \rho$ and $\mu(\bfx)=\mu$, where $1\leq \rho \leq n$ and $0\leq \mu \leq \rho-2$. Assume the vector $\bfx'$ has $p$ runs of ones of length $h_1,\ldots,h_p$. Then, first we have that
\begin{equation}\label{eq:1}
\sum_{i=1}^ph_i=w_H(\bfx') = \rho-1.
\end{equation}
Every run of ones of length $h_i$ in $\bfx'$ contributes $h_i-1$ pairs of two consecutive ones. Therefore,
\begin{equation}\label{eq:2}
\sum_{i=1}^p(h_i-1)=\mu.
\end{equation}
Together, from~(\ref{eq:1}) and~(\ref{eq:2}), we conclude that $p=\rho-\mu-1$. Furthermore, the number of solutions to~(\ref{eq:1}) (or~(\ref{eq:2})) is $\binom{\mu+p-1}{\mu}=\binom{\rho-2}{\mu}$. For every solution $h_1,\ldots,h_{\rho-\mu-1}$, let $k_0,k_1,\ldots,k_{\rho-\mu-1}$ be the number of zeros between the blocks of ones in $\bfx'$, where $k_0\geq 0, k_1,\ldots,k_{\rho-\mu-2}\geq 1$, and $k_{\rho-\mu-1}\geq 0$. Note that their sum is $n-1-(\rho-1) = n-\rho$, and thus, under the above constraints, the number of solutions to
$$\sum_{j=0}^{\rho-\mu-1}k_j=n-\rho$$ is $\binom{n-\rho+1}{\rho-\mu-1}$.

Finally, the number of options to choose the vector $\bfx'$ is the number of solutions to choose the runs of ones $h_1,\ldots,h_{\rho-\mu-1}$ and runs of zeros $k_0,\ldots,k_{\rho-\mu-1}$. Every choice of the vector $\bfx'$ determines the vector $\bfx$ up to choosing whether it starts with zero or one. Therefore, we get
$$N_n(\rho,\mu)=2\binom{\rho-2}{\mu}\binom{n-\rho+1}{\rho-\mu-1}.$$
\end{IEEEproof}

Next, the main result in this section is proved.
\begin{theorem}\label{thm:transversal}
The vector $\bfw=(w_{\bfx})_{\bfx\in \{0,1\}^{n-1}}$ defined by
\begin{displaymath}
w_{\bfx} 
= \left\{ \begin{array}{ll}
\frac{1}{\rho(\bfx)} & \textrm{if $\mu(\bfx)\leq 1$}\\
\frac{1}{\rho(\bfx)}\left(1-\frac{\mu(\bfx)}{\rho(\bfx)^2}\right) & \textrm{otherwise}
\end{array} \right.
\end{displaymath}
is a fractional transversal.
\end{theorem}
\begin{IEEEproof}
Let $\bfx$ be a length-$n$ binary vector with $\rho$ runs and $\mu$ middle-1-runs. We need to show that $\sum_{\bfy\in B_{D,1}(\bfx)}w_{\bfy} \geq 1$. It can be verified that this claim holds for $\rho=1,2,3$ or $\mu=0,1$ and thus we assume for the rest of the proof that $\rho\geq 4$ and $\mu\geq 2$. Note that for a fixed $\rho$, $w_{\bfx}$ is decreases when $\mu$ increases.

If a vector $\bfy\in B_{D,1}(\bfx)$ is received by deleting a middle-1-run bit then $\rho(\bfy)=\rho-2$ and $\mu-3\leq \mu(\bfy)\leq \mu-1$. Otherwise, $\rho(\bfy) = \rho$ and $\mu(\bfy)\leq \mu+1$ or, if it is the first or last bit which is a single-bit run, $\rho(\bfy) = \rho-1$ and $\mu(\bfy)\leq \mu$, however, the worst case in terms of the value of $w_{\bfy}$ is achieved for $\rho(\bfy) = \rho$ and $\mu(\bfy)= \mu+1$. Therefore, \vspace{-1ex}

\begin{small}
\begin{align*}
&\hspace{-2ex} \sum_{\bfy\in B_{D,1}(\bfx)}w_{\bfy} \geq \frac{\mu}{\rho-2}\left(1-\frac{\mu-1}{(\rho-2)^2}\right) + \frac{(\rho-\mu)}{\rho}\left(1-\frac{\mu+1}{\rho^2}\right) & \\
& \hspace{-1ex} = 1+\frac{2\mu}{\rho(\rho-2)}-\frac{\mu(\mu-1)}{(\rho-2)^3}-\frac{\mu+1}{\rho^2}+\frac{\mu(\mu+1)}{\rho^3},
\end{align*}
\end{small}
and thus it is enough to show that for $\rho\geq 4,2\leq \mu\leq \rho-2$,
\begin{small}
$$\frac{2\mu}{\rho(\rho-2)}-\frac{\mu(\mu-1)}{(\rho-2)^3}-\frac{\mu+1}{\rho^2}+\frac{\mu(\mu+1)}{\rho^3} \geq 0,$$
\end{small}
or
\begin{small}
$$\hspace{-2ex}\frac{2}{\rho(\rho-2)}\hspace{-0.3ex}-\hspace{-0.3ex}\frac{1}{\rho^2}\hspace{-0.5ex}\geq\hspace{-0.5ex} \frac{\mu-1}{(\rho-2)^3}\hspace{-0.5ex}+\hspace{-0.5ex}\frac{1}{\mu \rho^2}\hspace{-0.3ex}-\hspace{-0.3ex}\frac{\mu+1}{\rho^3}.$$
\end{small}
The function $$f(\mu)=  \frac{\mu-1}{(\rho-2)^3}+\frac{1}{\mu \rho^2}-\frac{\mu+1}{\rho^3}$$ in the range $2\leq \mu \leq \rho-2$ is maximized either when $\mu=2$ or $\mu=\rho-2$ and thus we need to show that\vspace{-1ex}
\begin{small}
$$\frac{2}{\rho(\rho-2)}-\frac{1}{\rho^2}\geq \frac{1}{(\rho-2)^3} + \frac{1}{2\rho^2}-\frac{3}{\rho^3},$$
\end{small}
and
\begin{small}
$$\frac{2}{\rho(\rho-2)}-\frac{1}{\rho^2}\geq \frac{\rho-3}{(\rho-2)^3} + \frac{1}{(\rho-2)\rho^2}-\frac{\rho-1}{\rho^3},$$
\end{small}
which holds for all $\rho\geq 4$.
\end{IEEEproof}

For a vector $\bfx$ with $\rho$ runs and $\mu$ middle-1-runs, we denote its weight by $w(\rho,\mu)$, as specified in Theorem~\ref{thm:transversal}. From Lemma~\ref{lem:count} and Theorem~\ref{thm:transversal} we conclude with the following upper bound on $\tau^*(\cH(\cG_D,1))$.
\begin{theorem}\label{thm: D new bound}
The value $\tau^*(\cH(\cG_D,1))$ satisfies
$$\tau^*(\cH(\cG_D,1)) \leq 2+\sum_{\rho=2}^{n-1}\sum_{\mu=0}^{\rho-2} N_{n-1}(\rho,\mu)w(\rho,\mu).$$
\end{theorem}
\begin{IEEEproof}
We calculate the upper bound on $\tau^*(\cH(\cG_D,1))$ according to the fractional transversal from Theorem~\ref{thm:transversal}, $\bfw =(w_{\bfx})_{\bfx\in\{0,1\}^{n-1}}$. Every vector $\bfx$ is assigned with a weight $w_{\bfx}=w(\rho,\mu)$ according to its number of runs $\rho$ and number of middle-1-runs $\mu$. Thus, we get this upper bound to be
$$\sum_{\bfx\in\{0,1\}^{n-1}} w_{\bfx} = 2+\sum_{\rho=2}^{n-1}\sum_{\mu=0}^{\rho-2} N_{n-1}(\rho,\mu)w(\rho,\mu).$$
\end{IEEEproof}

Table~\ref{table:deletion} summarizes the results of the different bounds discussed in this subsection. MB corresponds to the equivalent of the monotonicity upper bound, which is the value $\frac{2^n-2}{n-1}$ from~\cite{KK12}. ASPV corresponds to the average sphere packing value $\frac{2^n}{n+1}$ from Lemma~\ref{lem:D ASPV}. The third column is our upper bound results from Theorem~\ref{thm: D new bound}. The column titled GSPB~\cite{KK12} is the exact value of $\tau^*(\cH(\cG_D,1))$ from~(\ref{eq:GSPB deletion}), which this linear programming problem was numerically solved in~\cite{KK12} for $n\leq 14$. Since this linear programming has a large number of constraints and variables it is numerically hard to solve it for larger values of $n$. The last column LB corresponds to the lower bound, which is the best known construction of single-deletion codes from~\cite{VT65}.
\begin{table}[h]
\begin{center}
\caption{Deletion Channel Comparison}\label{table:deletion}
\begin{tabular}{|c||c|c|c|c|c|}
  \hline
  $n$ & MB~\cite{KK12} & ASPV & Theorem~\ref{thm: D new bound} & GSPB~\cite{KK12} & LB~\cite{VT65} \\ \hline \hline
  5 & 7       & 5      & 7      & 6     & 6     \\
  6 & 12      & 9      & 12     & 10    & 10    \\
  7 & 21      & 16     & 20     & 17    & 16    \\
  8 & 36      & 28     & 35     & 30    & 30    \\
  9 & 63      & 51     & 61     & 53    & 52    \\
  10 & 113    & 93     & 109    & 96    & 94    \\
  11 & 204    & 170    & 197    & 175   & 172   \\
  12 & 372    & 315    & 358    & 321   & 316   \\
  13 & 682    & 585    & 657    & 593   & 586   \\
  14 & 1260   & 1092   & 1212   & 1104  & 1096  \\
  15 & 2340   & 2048   & 2251   & ?     & 2048  \\
  16 & 4368   & 3855   & 4202   & ?     & 3856  \\
  17 & 8191   & 7281   & 7882   & ?     & 7286  \\
  18 & 15420  & 13797  & 14845  & ?     & 13798 \\
  19 & 29127  & 26214  & 28059  & ?     & 26216 \\
  20 & 55188  & 49932  & 53202  & ?     & 49940 \\
  21 & 104857 & 95325  & 101163 & ?     & 95326 \\
  22 & 199728 & 182361 & 192850 & ?     & 182362\\
  23 & 381300 & 349525 & 368478 & ?     & 349536\\
  \hline
\end{tabular}
\end{center}
\end{table}

\subsection{Grain Errors}
The grain-error channel is a recent model which was studied mainly for granular media with applications to magnetic recording technologies~\cite{WNP97,WWKM09}. In this medium, the information is stored in individual grains which their magnetization can hold a single bit of data. However, since the size of these grains is very small, the information bits are written to the grains without knowing in advance their exact location~\cite{MBK11}. Typically, the bit cell is larger than a single-grain and in this case the polarity of a cell is determined by the last bit that was written into it. This kind of errors is called \textbf{\emph{grain-errors}}. We will follow the model studied by previous works which assume that the first bit smears its adjacent one to the right. There are several recent studied of this model which analyzed its information theory behavior~\cite{ISW11}, proposed code constructions, and upper bounds~\cite{GYD13,GYD13b,KZ13,MBK11,SR11,SR13}.

The grain-error channel is very similar to the deletion channel, however in this case the length of the received words remains the same.
The graph describing this channel model is $\cG_G=(X_G,E_G)$, where $X_G = \{0,1\}^n$ and
\begin{align*}
E_G = \{(\bfx,\bfy) : & \ \bfx,\bfy \in \{0,1\}^n, \textmd{ and there exists } 2\leq i\leq n & \\
                      & \textmd{ such that }\bfy =\bfx +\bfe_i\textmd{ and }x_i\neq x_{i-1}\}. &
\end{align*}
The radius one ball for some $\bfx\in\{0,1\}^n$ is the set $B_{G,1}(\bfx) = \{\bfy \in \{0,1\}^n \ : \ (\bfx,\bfy)\in E_G\}$. The hypergraph for the single grain-error channel becomes $\cH(\cG_G,1) = (X_{G,1},\cE_{G,1})$, where $X_{G,1} = \{0,1\}^n$ and $\cE_{G,1} = \{B_{G,1}(\bfx) \ : \ \bfx\in \{0,1\}^n \} $. Finally, the generalized sphere packing bound for the single grain-error channel is
\begin{equation}
\hspace{-0.5ex}\tau^*(\cH(\cG_{G},1))\hspace{-0.5ex}=\hspace{-0.5ex}\min\hspace{-0.5ex} \bigg\{\hspace{-0.6ex}\sum_{\bfx\in\{0,1\}^{n}}\hspace{-1.5ex}w_{\bfx}  : \hspace{-1.2ex} \sum_{\bfy\in B_{G,1}(\bfx)}\hspace{-2ex}w_{\bfy} \geq 1, \forall
\bfx\in\{0,1\}^n \bigg\}.
\end{equation}

The size of the ball $B_{G,1}(\bfx)$ can be given by $\deg_{G,1}(\bfx) = \rho(\bfx)$, where, as before, $\rho(\bfx)$ is the number of runs in $\bfx$. It is also verified that if $\bfy\in B_{G,1}(\bfx)$ then $\rho(\bfy)\leq \rho(\bfx)$ and thus the graph $\cG_G$ satisfies the monotonicity property. These results were verified both in~\cite{GYD13} and~\cite{KZ13} and showed that the vector $\bfw = (w_\bfx)_{\bfx\in\{0,1\}^n}$ given by $w_{\bfx}=\frac{1}{\rho(\bfx)}$, is a fractional transversal. Accordingly, the corresponding upper bound, called here the monotonicity upper bound, on $\tau^*(\cH(\cG_{G},1))$ becomes
$$MB(\cG_G,1) =\frac{2^{n+1}-2}{n}.$$
This bound is slightly improved in~\cite{GYD13b} by noticing that if there is a code with odd number of codewords, then there exists a code with one more codeword, and thus this upper bound becomes $2\lfloor\frac{2^{n+1}-2}{2n}\rfloor$.

The average sphere packing value in this case is calculated in the next lemma.
\begin{lemma}\label{lem:G ASPV}
The average sphere packing value of the graph $\cG_{G}$ for $r=1$ is
$$ASPV(\cG_{G},1) = \frac{2^{n+1}}{n+1}.$$
\end{lemma}
\begin{IEEEproof}
The size of the radius one ball centered in $\bfx\in\{0,1\}^n$ is $\deg_{G,1}(\bfx) = \rho(\bfx)$. Thus, the average size of a ball is
\begin{align*}
& \frac{1}{2^n}\sum_{\bfx\in\{0,1\}^n} \deg_{G,1}(\bfx) = \frac{1}{2^n}\sum_{\rho=1}^n 2\binom{n-1}{\rho-1} \rho & \\
& = \frac{1}{2^{n-1}}\sum_{\rho=0}^{n-1} \binom{n-1}{\rho} (\rho+1) = \frac{1}{2^{n-1}}(2^{n-1}+(n-1)2^{n-2})& \\
& = \frac{n+1}{2}. &
\end{align*}
Thus, the average sphere packing value becomes
$$\frac{2^{n}}{(n+1)/2} = \frac{2^{n+1}}{n+1}.$$
\end{IEEEproof}

Note that very similarly to the deletion channel, the fractional transversal given by the monotonicity property is suboptimal. We carry similar steps as in the previous subsection in order to give a better fractional transversal, stated in the next theorem.
\begin{theorem}\label{thm:transversal grain}
The vector $\bfw=(w_{\bfx})_{\bfx\in \{0,1\}^{n}}$ defined by
\begin{displaymath}
w_{\bfx}
= \left\{ \begin{array}{ll}
\frac{1}{\rho(\bfx)} & \textrm{if $\mu(\bfx)\leq 1$}\\
\frac{1}{\rho(\bfx)}\left(1-\frac{\mu(\bfx)}{\rho(\bfx)^2}\right) & \textrm{otherwise}
\end{array} \right.
\end{displaymath}
is a fractional transversal.
\end{theorem}
\begin{IEEEproof}
Let $\bfx$ be a binary vector of length $n$ with $\rho$ runs and $\mu$ middle-1-runs. We will show that $\sum_{\bfy\in B_{G,1}(\bfx)}w_{\bfy} \geq 1$. As in the proof of Theorem~\ref{thm:transversal}, it is possible to verify that this property holds for $\rho=1,2,3$ and $\mu=0,1$, so we assume that $\rho\geq 4$ and $\mu \geq 2$.

If a vector $\bfy\in B_{G,1}(\bfx)$ is received by a single grain-error of a middle-1-run bit then $\rho(\bfy)=\rho-2$ and $\mu(\bfy)\leq \mu-1$. Otherwise, $\rho(\bfy) = \rho$ and $\mu(\bfy)\leq \mu+1$ or, in case the last bit errs, $\rho(\bfy) = \rho-1$ and $\mu(\bfy)\leq \mu$, however, the worst case is achieved for $\rho(\bfy) = \rho$ and $\mu(\bfy) = \mu+1$. Hence, we get

\vspace{-2ex}
\begin{small}
\begin{align*}
\hspace{-2ex} \sum_{\bfy\in B_{G,1}(\bfx)}w_{\bfy} \geq & \frac{1}{\rho}\left(1-\frac{\mu}{\rho^2}\right) + \frac{\mu}{\rho-2}\left(1-\frac{\mu-1}{(\rho-2)^2}\right) & \\
& + \frac{(\rho-\mu-1)}{\rho}\left(1-\frac{\mu+1}{\rho^2}\right) & \\
& \hspace{-1ex} \geq 1+\frac{2\mu}{\rho(\rho-2)}-\frac{\mu(\mu-1)}{(\rho-2)^3}-\frac{\mu+1}{\rho^2}+\frac{\mu(\mu+1)}{\rho^3}.\vspace{-2ex}
\end{align*}
\end{small}
The rest of the proof is identical to the proof of Theorem~\ref{thm:transversal}.
\end{IEEEproof}

Finally, we conclude with the following theorem.
\begin{theorem}\label{thm: G new bound}
The value $\tau^*(\cH(\cG_G,1))$ satisfies
$$\tau^*(\cH(\cG_G,1)) \leq 2+\sum_{\rho=2}^{n}\sum_{\mu=0}^{\rho-2} N_{n}(\rho,\mu)w(\rho,\mu).$$
\end{theorem}

Table~\ref{table:grain} summarizes the improvements and results discussed in this section on the cardinalities of single-grain error-correcting codes.
In the last column we gave the cardinalities of the best known to us codes taken from~\cite{GYD13b},~\cite{SR11}, and~\cite{SR13}. For $5\leq n\leq 8$ the lower bound coincides with the best known upper bound~\cite{SR11}. For $n=9,10,11$ the best known upper bound is $88,176,352$, respectively~\cite{SR13}, while for $n\geq 12$ the best known upper bound is the monotonicity upper bound from~\cite{GYD13b,KZ13}.
\begin{table}[h]
\begin{center}
\caption{Grain-Error Channel Comparison}\label{table:grain}
\begin{tabular}{|c||c|c|c|c|}
  \hline
  $n$& MB~\cite{GYD13b,KZ13}     & ASPV   & Theorem~\ref{thm: G new bound} & LB \\ \hline \hline
  5  & 12     & 10     & 12                             & 8~\cite{SR11}\\
  6  & 20     & 18     & 20                             & 16~\cite{SR11}\\
  7  & 36     & 32     & 35                             & 26~\cite{SR11}\\
  8  & 62     & 56     & 60                             & 44~\cite{SR11}\\
  9  & 112    & 102    & 108                            & 72~\cite{SR13} \\
  10 & 204    & 186    & 196                            & 112~\cite{SR13}\\
  11 & 372    & 341    & 358                            & 210~\cite{GYD13b}\\
  12 & 682    & 630    & 656                            & 372~\cite{SR13}\\
  13 & 1260   & 1170   & 1212                           & 702~\cite{GYD13b} \\
  14 & 2340   & 2184   & 2250                           & 1272~\cite{SR13}\\
  15 & 4368   & 4096   & 4202                           & 2400~\cite{GYD13b}\\
  16 & 8190   & 7710   & 7882                           & 4522~\cite{SR13}\\
  17 & 15420  & 14563  & 14844                          & 8428~\cite{SR13}\\
  18 & 29126  & 27594  & 28058                          & 15348~\cite{GYD13b}\\
  19 & 55188  & 52428  & 53202                          & 27596~\cite{GYD13b}\\
  20 & 104856 & 99864  & 101162                         & 52432~\cite{GYD13b}\\
  21 & 199728 & 190650 & 192850                         & 99880~\cite{GYD13b}\\
  22 & 381300 & 364722 & 368478                         & 190652~\cite{GYD13b}\\
  23 & 729444 & 699050 & 705510                         & 364724~\cite{GYD13b}\\
  \hline
\end{tabular}
\end{center}
\end{table}

\section{Projective Spaces}\label{sec:projective}
In this section, we explain an example where there is no monotonocity property, yet we benefit from the graph automorphisms and we simplify the linear programming again.

Koetter and Kschischang ~\cite{KK08} modeled codes as subsets of projective space $\mathbb{F}_q^n$, the set of linear subspaces of $\mathbb{F}_q^n$, or of Grassmann space $\cG(n,k)$, the subset of linear subspaces of $\mathbb{F}_q^n$ having dimension $k$. Subsets of $\mathbb{F}_q^n$ are called \emph{projective codes} and similar to previous sections, it is desired to select elements with large distance from each other.

Let us first introduce the graph $\cG_P=(X_P,E_P)$ for projective codes, where $X_P$ is the set of all linear subspaces in $\mathbb{F}_q^n$ and
\begin{align*}
E_P=\hspace{-.4ex}\{\{x,y\}\hspace{-.4ex}:\hspace{-.4ex} x \hspace{-.4ex}\subset\hspace{-.4ex} y \text{ or } y\hspace{-.4ex}\subset\hspace{-.4ex} x,\text{ and }|\hspace{-.4ex}\dim(x)-\dim(y)\hspace{-.2ex}|=1\},
\end{align*}
and using the path distance $d_P(x,y)$ defined on graph $\cG_P$ we define
\begin{align*}
\cB_{P,r}(x)=\{y\in \cX_P: d_P(x,y) \leq r\}.
\end{align*}
The corresponding hypergraph is $\cH(\cG_P,r)=(X_{P,r},\cE_{P,r})$, such that $X_{P,r}=X_P$ and $\cE_{P,r}=\{ \cB_{P,r}(x) : x\in X_P\}$. The generalized sphere packing bound becomes
\vspace{-1ex}
\begin{align*}
&\tau^*\hspace{-0.1ex}(\cH(\cG_P,r\hspace{-0.1ex})\hspace{-0.1ex})\hspace{-0.4ex}=\hspace{-0.4ex}\min\bigg\{\hspace{-.8ex}\sum_{x\in X_P} \hspace{-1ex}w(x)\hspace{-0.5ex} :\hspace{-0.5ex} \forall  x\in \hspace{-0.6ex}\cX_P,  \hspace{-2.5ex}\sum_{y\in \cB_{P,r}(x)}\hspace{-2ex}w_y\hspace{-0.4ex}\geq\hspace{-0.4ex} 1, w_x\hspace{-0.4ex}\geq\hspace{-0.4ex} 0\hspace{-0.3ex}   \bigg\}.
\end{align*}

Assume $x_1$ and $x_2$ are elements in $X_P$ with same dimension $k$. There exist an injective linear transform $\cT: \mathbb{F}_q^n \rightarrow \mathbb{F}_q^n$ mapping the basis of $x_1$ into a basis for $x_2$. Note that $x\subset y$ if and only if $\cT(x)\hspace{-.4ex}\subset\hspace{-.4ex} \cT(y)$. Hence, all such linear transforms are automorphisms on $\cG_P$, which means for any $x_1,x_2\in X_P$ of the same dimensions, there exist an automorphism mapping between them. Therefore, they lie in a same equivalence class. So we assign a same transversal weight to all the subspaces with the same dimension. We also need to find the size and the distribution of elements in $\cB_{P,r}(x)$. The general formula is given in ~\cite{TV11} but we only study the case $r=1$. Given $x$ with dimension $k$ in $X_P$, there are $\Gb{k}{k-1}=2^k-1$ subspaces of dimension $k-1$ in $\cB_{P,1}(x)$, where
\vspace{-1ex}
$$\Gb{n}{m}={(2^n-1)(2^{n-1}-1)\cdots(2^{n-m+1})\over (2^m-1)(2^{m-1}-1)\cdots (2^1-1) }$$
is the number of subspaces of dimension $m$ in a space of dimension $n$. There are also ${2^n-2^k\over 2^k}=2^{n-k}-1$ subspaces of dimension $k+1$ in $\cB_{P,1}(x)$ that include $x$. Therefore, there are $(2^k-1)+(2^{n-k}-1)+1$ elements in $\cB_{P,1}(x)$. So,
\vspace{-1ex}
\begin{align}
&\tau^*(\cH(\cG_P,r))=\min\bigg\{\sum_{k=0}^n w_k\Gb{n}{k}:\forall\; 0\leq k\leq n\label{eq: GSPB-Projective}\\
&w_k+(2^k-1)w_{k-1}+(2^{n-k}-1)w_{k+1}\geq 1,w_k\geq 0\bigg\}. \nonumber
\end{align}

It is shown that there exist automorphisms which map a fixed subspace of dimension $k$ to a fixed subspace of dimension $n-k$ (see ~\cite{BEV}.) So, subspaces of dimension $k$ and $n-k$ are also in same equivalence classes and we assign same weights to them. Also note that  $\Gb{n}{k}=\Gb{n}{n-k}$. Hence we benefit from a very nice symmetry and we set $w_k=w_{n-k}$ to halve both the number of constraints and the parameters in the linear programming. Optimal transversal weights for $n\leq 11$ are listed in Table \ref{table:PCodes}.
\begin{table}[h]\label{table: Projective weight}
\begin{center}
\caption{Projective codes: upper bounds and weights for $r=1$}\label{table:PCodes}
\begin{tabular}{|c||c|c|c|c|}
  \hline
  $n$ & $w_0^*,w_1^*,\cdots,w_{\floor{n\over 2}}^*$   & ASPV & GSPB  & \cite{BVP}  \\ \hline \hline
  2   &  1, 0   & 1      & 1      & - \\
  3   & 1, 0     & 3     & 2     & - \\
  4   & 0.83, 0.17, 0    & 8     & 6     & 6 \\
  5   & 0.67, 0.34, 0     & 30     & 22     & 20 \\
  6   &  0, 0.30, 0.07, 0   & 159     & 132     & 124 \\
  7  & 0, 0.29, 0.15, 0    & 1142    & 834    & 776 \\
  8  & 1, 0, 0.14, 0.03, 0     & 11364    & 9460    & 9268 \\
  9  & 1, 0, 0.13, 0.07, 0    & 157860    & 116656    & 107419 \\
  10  & 1, 0, 0, 0.066, 0.016, 0   & 3073031   & 2566390   & - \\
  11  & 1, 0, 0, 0.065, 0.032, 0   & 84047153   & 62462160   & - \\ \hline
\end{tabular}
\end{center}
\end{table}

It is interesting to see that $w_{\floor{n\over 2}}=0$ for all $n>2$, which is not surprising since $\Gb{n}{\floor{n\over 2}}$ is the largest coefficient in the cost function. This leads us to a greedy approach of starting from the middle, which has the highest impact on cost function; minimizing it, i.e.  $w_{\floor{n\over 2}}=0$; and then moving toward the tails where we pick the least possible value to satisfy the constraints. We call it as the {\textbf{\textit{greedy}}} weight assignment, which is expressed as
\vspace{-1ex}
\begin{align}
&w_{[{n\over 2}]}^*=0, \;\; \text{ and for all $k$:  }\;\; 0\leq k<\floor{n\over 2}, \nonumber\\
&w_k^*=\max\{{{1-w_{k+1}^*-(2^{n-k-1}-1)w_{k+2}^*}\over 2^{k+1}-1},0\},\label{eq:greedyP}\\
&w_k^*=w_{n-k}^*, \text{ and if} \; w_0^*=w_1^*=0, {\text{ then }} w_0^*=1.\nonumber
\end{align}
It is clear that the greedy output has the transversal property and lies in the feasible set. In fact, $w^*$ for $k<\floor{\frac{n}{2}}$ is given by
\vspace{-1ex}
\begin{displaymath}
w_k^*
= \left\{ \begin{array}{ll}
\frac{1}{2^{k+1}-1} & \textrm{if $k\equiv \floor{n/2}-1 $ mod $4$}\\
\frac{2}{2^{k+2}-1} & \textrm{if $k\equiv \floor{n/2}-2 $ mod $4$}\\
0 & \textrm{otherwise,}
\end{array} \right.
\end{displaymath}
\vspace{-1ex}
with the only exception of

\begin{displaymath}
w_k^*
= \left\{ \begin{array}{ll}
\frac{1}{2(2^{k+1}-1)} & \textrm{if $k=\frac{n}{2}-1$}\\
\frac{2^{k+3}-3}{(2^{k+1}-1)(2^{k+2}-2)} & \textrm{if $k=\frac{n}{2}-2$}\\
\end{array} \right.
\end{displaymath}
for $n$ even. The following theorem also shows the optimality of greedy assignment in our scheme (See appendix ~\ref{app:projective optimality} for the proof.)
\vspace{-1.5ex}
\begin{theorem}\label{thm:projectiveoptimal} Let $\cG_P$ be the associated graph with projective code when $\mathbb{F}_q^n$ is the space and $\bf{w^*}$ be defined as (\ref{eq:greedyP}), then
\vspace{-1.5ex}
\begin{align*}
\tau^*(\cH(\cG_P,r))=\sum_{k=0}^n w_k^*\Gb{n}{k}.
\end{align*}
\end{theorem}

\section{Conclusions and Discussion}\label{sec:conclusion}
In this paper we presented a generalization of the sphere packing bound, based upon a recent work by Kulkani and Kiyavash for deriving upper bounds on the cardinality of deletion-correcting codes. Our scheme can provide upper bound on the cardinality of codes according to any error channel. The main challenge in deriving this upper bound is the solution of a linear programming problem, which in many cases is not easy to find. We found this solution for the $Z$ channel and projective spaces in case of radius one. In the other setups studied here, namely the limited magnitude, deletion, and grain-error channels, we didn't completely solve the linear programming problem but found a corresponding upper bound, which is a valid upper bound on the codes cardinalities in each case. Thus, solving the linear programming, in order to find the generalized sphere packing bound for each error channel, still remains an interesting open problem. We also mention that other error channels can be studied as well using the scheme presented in the paper.

Lastly, we follow up on the question we asked in the Introduction about the validity of the average sphere packing value. Even though in general it is not a valid upper bound, we believe that there are some conditions under which this value will hold as an upper bound, and finding these conditions remains as open problem.

\appendices
\section{Optimal transversal weight for $Z$ channel}\label{app:Z optimality}

In this section, we go over lemmas and proofs that has been used in section ~\ref{sec:Z channel}.

\begin{lemma}\label{lemma: bound on D}
For all $r\in \mathbb{N}$, $|D_m|\leq (2r)^{m-r+1}$.
\end{lemma}
\begin{IEEEproof}
The proof is based on induction on $m$. Let us assume $|D_i|\leq (2r)^{i-r+1}$ for all $i\leq m-1$. Therefore,
\begin{align*}
D_m&=-({r!\over (r-1)!}D_{m-1}+{r!\over (r-2)!}D_{m-2}+\cdots+{r!\over 0!}D_{m-r})\\
&\leq {r!|D_{m-1}|\over (r-1)!}+{r!|D_{m-2}|\over (r-2)!}+\cdots+{r!|D_{m-r}|\over 0!}\\
&\leq {r!(2r)^{m-r}\over (r-1)!}+{r!(2r)^{m-r-1}\over (r-2)!}+\cdots+{r!(2r)^{m-2r+1}\over 0!}\\
&= (2r)^{m-r+1}({r!(2r)^{-1}\over (r-1)!}+{r!(2r)^{-2}\over (r-2)!}+\cdots+{r!(2r)^{-r}\over 0!})\\
&\leq (2r)^{m-r+1}(2^{-1}+2^{-2}+\cdots+2^{-r})\\
&\leq (2r)^{m-r+1}.
\end{align*}
\end{IEEEproof}

The idea behind the optimality proof is to write the cost function $f(\bf{w})$ as a non-negative linear combination of some other cost functions denoted by $f_i(\bf{w})$ and show $\bf{w}^*$ is the a feasible point which minimizes them all, and hence $\bf{w}^*$ also minimizes the cost function and is the desired optimal solution.

Let us define cost functions $f_i(\bf{w})$ for all $0\leq i \leq n$ as
\begin{align*}
&f_k({\bf{w}})=w_0&k=0,\\
&f_k({\bf{w}})=\sum_{i=0}^rw_{k+i}\binom{k+r}{r-i} &\forall 1\leq k\leq n-r,\\
&f_k({\bf{w}})=w_k  &\forall n-r<k\leq n.
\end{align*}

From ~(\ref{eq:Z GSPB2}), any feasible $w$ should satisfy
\begin{align*}
&f_k({\bf{w}})\geq 1&k=0,\\
&f_k({\bf{w}})\geq 1 &\forall 1\leq k\leq n-r,\\
&f_k({\bf{w}})\geq 0 &\forall n-r<k\leq n;
\end{align*}
And, ${\bf{w}}={\bf{w}}^*$ gives us equalities in all of them. We are required to show ${\bf{w}}^*$ also minimizes $f({\bf{w}})$, where
\begin{align*}
f({\bf w})= \sum_{i=0}^nw_{i}\binom{n}{i}.
\end{align*}

In order to prove the optimality, we show $f({\bf w})$ can be written as
\begin{align*}
f({\bf w})=y_0f_0({\bf w})+y_1f_1({\bf w})+\cdots+y_nf_n({\bf w}),
\end{align*}
where $y_i$'s are some non-negative constants. Hence, for any transversal weight ${\bf w}$ we have
\begin{align*}
f({\bf w})&\geq y_0f_0({\bf{w}}^*)+y_1f_1({\bf{w}}^*)+\cdots+y_nf_n({\bf{w}}^*)\\
&=y_0+y_1+\cdots+y_{n-r}=f(w^*).
\end{align*}

We first show the choice of ${\bf{y}}$ is unique. Then the problem reduces to show the non-negativity of ${\bf{y}}$. Note that the cost functions $f_i({\bf{w}})$ are inner products of ${\bf{w}}$ with some non-negative vectors ${\bf{m}}_i$, i.e. $f_i({\bf{w}})=<{\bf{m}}_i,{\bf{w}}>$, where
\begin{align*}
    m_{ij}=
    \begin{cases}
      1, & \text{if}\ i=j=0 \\
      1, & \text{if}\ i=j, \text{and } n-r< i\leq n\\
      \binom{i+r}{j}, & \text{if}\ 1\leq i\leq n-r, \text{and } i\leq j\leq i+r\\
      0, & \text{otherwise}.
    \end{cases}
\end{align*}

Now, if we form the $(n+1)\times (n+1)$ matrix $M$ with elements $\{m_{ij}\}$, the problem of finding ${\bf{y}}$ will be equivalent to solving $M^T{\bf{y}}={\bf{c}}$, where ${\bf{c}}=(\binom{n}{0},\binom{n}{1},\cdots,\binom{n}{n})$. $M$ is an upper triangular matrix with non-zero elements on the main diagonal, and hence $M$ is non-singular and the solution is unique. Since $M$ is upper triangular, $M^Ty=c$ gives us the following recursions on $y_i$'s:
\begin{align}\label{eq:y recursive}
&y_0=0,\nonumber\\
&y_i={1\over \binom{r+i}{i}}(\binom{n}{i}-\sum_{j=max\{i-r,1\}}^{i-1}\binom{r+j}{i}y_{j})\\
&\forall i:\; 1\leq i\leq n-r,\nonumber
\end{align}
and
\begin{align*}
&y_i=\binom{n}{i}-\sum_{j=max\{i-r,1\}}^{n-r}\binom{r+j}{i}y_{j}.\\
&\forall i:\; n-r< i\leq n.
\end{align*}

Lemma ~\ref{lemma:y explicit} gives an explicit formula for $y_k$ when $0\leq k\leq n-r$, which again uses the sequence $D_i$ defined in ~(\ref{eq:D recursive}). We put the proof at the end.

\begin{lemma}\label{lemma:y explicit}
If ${\bf{y}}$ is the solution to $M^T{\bf{y}}={\bf{c}}$, then
\begin{align*}
&y_0=1, &\\
&y_k={r!\over (k+r)!}\sum_{m=1}^{k}{n!\over (n-m)!}D_{k-m+r-1} &\forall 1\leq k\leq n-r.
\end{align*}
\end{lemma}

Let us benefit from a simple change of variables to get
\begin{align*}
z_k&:=y_{n-k}{(n-k+r)!\over r!n!}=\sum_{m=1}^{n-k}{D_{n-k-m+r-1}\over (n-m)!}\\
&\;=\sum_{m=0}^{n-k-1}{D_{m+r-1}\over (m+k)!}={1\over k!}+\sum_{m=1}^{n-k-1}{D_{m+r-1}\over (m+k)!},
\end{align*}
which is proven to be positive if $k\geq 4r-1$ by using the same argument as $w_k^*>0$ when $k\geq 3r-1$ in ~(\ref{proof:Z 3k-1}). Also, we can verify $z_k\geq 0$ for the values of $r\leq k\leq 4r-2$ by finding some numbers $n'_k$ such that
\begin{align*}
&\sum_{m=k}^{n'_k}{D_{m+r-k-1}\over m!}\geq {1\over (2r)^{k}}(e^{2r}-\sum_{m=0}^{n_k}{(2r)^m\over m!}),
\end{align*}
and we get $z_k>0$ for all $n>n_k$ because
\begin{align*}
z_k&=\sum_{m=k}^{n-1}{D_{m+r-k-1}\over m!}\\
&=\sum_{m=k}^{n_k}{D_{m+r-k-1}\over m!}+\sum_{m=n_k+1}^{n-1}{D_{m+r-k-1}\over m!}\\
&\geq \sum_{m=k}^{n_k}{D_{m+r-k-1}\over m!}-\sum_{m=n_k+1}^{n-1}{(2r)^{m-k}\over m!}\\
&> \sum_{m=k}^{n_k}{D_{m+r-k-1}\over m!}-{1\over (2r)^{k}}\sum_{m=n_k+1}^{\infty}{(2r)^{m}\over m!}\\
&=\sum_{m=k}^{n_k}{D_{m+r-k-1}\over m!}-{e^{2r}-\sum_{m=0}^{n_k}{(2r)^m\over m!}\over (2r)^{k}}\geq 0.
\end{align*}
Finally, we check the values of $z_k$ for the finite set of $k<4r-1$ and $n\leq n'_k$.

In order to show $y_k\geq 0$ for $k>n-r$,  we rewrite the expression for $y_k$ when $k>n-r$ as
\begin{align*}
&y_k=\binom{n}{k}-\sum_{j=max\{k-r,1\}}^{n-r}\binom{r+j}{k}y_{j}=\\
&\binom{\hspace{-.2ex}n\hspace{-.2ex}}{k}\hspace{-.5ex}-y_{n-r}\binom{\hspace{-.2ex}n\hspace{-.2ex}}{k}\hspace{-.5ex}-y_{n-r-1}\binom{\hspace{-.2ex}n\hspace{-.2ex}-\hspace{-.2ex}1\hspace{-.2ex}}{k}-\hspace{-.1ex}\cdots-\hspace{-.5ex}y_{n-r-p}\binom{\hspace{-.2ex}n\hspace{-.2ex}-\hspace{-.2ex}p\hspace{-.2ex}}{k},
\end{align*}
where $n-r-p=max\{k-r,1\}$. So, $y_k\geq 0$ is equivalent to
\begin{align*}
&y_{n-r}\leq \\
&1-{\binom{n-1}{k}\over \binom{n}{k}}y_{n-r-1}-{\binom{n-2}{k}\over \binom{n}{k}}y_{n-r-2}-\cdots-{\binom{n-p}{k}\over \binom{n}{k}}y_{n-r-p}.
\end{align*}
By assuming $y_i\geq 0$ for all $i\leq n-r$ and the recursive formula for $y_k$'s, we have
\begin{align*}
y_{n-r}\hspace{-.4ex}=&1\hspace{-.4ex}-{\binom{n-1}{n-t}\over \binom{n}{n-r}}y_{n\hspace{-.1ex}-\hspace{-.1ex}r\hspace{-.1ex}-1}\hspace{-.2ex}-\hspace{-.1ex}{\binom{n-2}{n-r}\over \binom{n}{n-r}}y_{n\hspace{-.1ex}-\hspace{-.1ex}r\hspace{-.1ex}-\hspace{-.1ex}2}-\hspace{-.1ex}\cdots\hspace{-.1ex}-{\binom{n-q}{n-r}\over \binom{n}{n-r}}y_{n-r-q}\\
\leq &1\hspace{-.4ex}-{\binom{n-1}{n-t}\over \binom{n}{n-r}}y_{n\hspace{-.1ex}-\hspace{-.1ex}r\hspace{-.1ex}-\hspace{-.1ex}1}\hspace{-.2ex}-\hspace{-.1ex}{\binom{n-2}{n-r}\over \binom{n}{n-r}}y_{n\hspace{-.1ex}-\hspace{-.1ex}r\hspace{-.1ex}-\hspace{-.1ex}2}-\hspace{-.1ex}\cdots\hspace{-.1ex}-{\binom{n-p}{n-r}\over \binom{n}{n-r}}y_{n-r-p},
\end{align*}
where $n-r-q=max\{n-2r,1\}\leq max\{k-r,1\}=n-r-p$. Now, it suffices to show
\begin{align*}
&{\binom{n-\ell}{k}\over \binom{n}{k}}\leq {\binom{n-\ell}{n-r}\over \binom{n}{n-r}}\Longleftrightarrow {(n-k)!\over (n-\ell-k)!}\leq{r!\over (r-\ell)!}\\
\Longleftrightarrow& (n\hspace{-.2ex}-\hspace{-.2ex}k)(n\hspace{-.2ex}-\hspace{-.2ex}k\hspace{-.2ex}-\hspace{-.2ex}1)\cdots (n\hspace{-.2ex}-\hspace{-.2ex}k\hspace{-.2ex}-\hspace{-.2ex}\ell)\leq r(r\hspace{-.2ex}-\hspace{-.2ex}1)\cdots (r\hspace{-.2ex}-\hspace{-.2ex}\ell),
\end{align*}
which always holds since $n-k<r$. In short, for any fixed radius, one can verify the optimality in a very same fashion as feasibility by just proving $z_k\geq 0$ for all $r\leq k \leq 4r-2$ and the non-negativity of the remaining $y_k$'s follows immediately. Doing so, we proved the optimality of our transversal weight assignment for all $r\leq 20$.
\begin{IEEEproof}[Proof of Lemma ~\ref{lemma:y explicit}]
We define sequences $\{y_i^j(n)\}$ for $1\leq j \leq n-r$ as
\begin{align*}
&y_i^j(n)=0, &1\leq i<j,\\
&y_i^j(n)=\binom{n}{j}/\binom{r+j}{j}={n!r!\over (r+j)!(n-j)!} ,&i=j,\\
&y_i^j(n)=-\sum_{\ell=max\{i-r,1\}}^{i-1}{\binom{r+\ell}{i}\over \binom{r+i}{i}}y_{\ell}^j(n)  &j< i\leq n-r;
\end{align*}
And, define $y'_i(n)$ as $y'_i(n)=\sum_{j=1}^{n-r}y_i^j(n)$. It is easy to see
\begin{align*}
&y'_i(n)={1\over \binom{r+i}{i}}(\binom{n}{i}-\sum_{\ell=max\{i-r,1\}}^{i-1}\binom{r+\ell}{i}y'_{j}(n))\\
&\forall i:\; 1\leq i\leq n-r.
\end{align*}
So, $y'_i$ satisfies the same recursive relation as ~(\ref{eq:y recursive}) and hence it is nothing but $y_i$. On the other hand, using the change of variables $\delta_i^j(n)=y_i^j(n)(r+i)!{(n-j)!\over n!r!}$ gives us
\begin{align*}
&\delta_i^j(n)=0, & 1\leq i<j,\\
&\delta_i^j(n)=1, & i=j,\\
&{\delta_i^j(n)\over r!}+{\delta_{i-1}^j(n)\over (r-1)!}+\cdots+{\delta_{i-r}^j(n)\over 0!}=0, & j<i\leq n-r.
\end{align*}
So, $\delta_i^j(n)$ is nothing but $D_{i-j+r-1}$ defined in ~(\ref{eq:D recursive}). Substituting this value in the formula for $y'$ gives
\begin{align*}
&y_i=\sum_{j=1}^{n-r}y_i^j(n)=\sum_{j=1}^{n-r}\delta_i^j(n){n!r!\over (r+i)!(n-j)!}\\
&={r!\over (r+i)!}\sum_{j=1}^{n-r}{n!\over (n-j)!}D_{i-j+r-1}\\
&={r!\over (r+i)!}\sum_{j=1}^{i}{n!\over (n-j)!}D_{i-j+r-1}.
\end{align*}
\end{IEEEproof}

\section{Optimal transversal weight for projective codes}\label{app:projective optimality}

The idea behind the proof is to again write the cost function $\sum_{k=0}^nw_k\Gq{n}{k}$ as a non-negative linear combination of some other cost functions $f_k({\bf w})$, where ${\bf w}^*$ minimizes them all and so minimizes the cost function. We strongly benefit from the symmetry in the optimal transversal i.e. $w_k^*=w_{n-k}^*$ and we discuss the proof for all indices $k\leq \floor{\frac{n}{2}}$ without loss of generality. Let us define the partial cost functions $f_k(w)$ as
\begin{displaymath}
f_k({\bf w})
= \left\{ \begin{array}{ll}
w_{k+1}+(2^{k+1}-1)w_{k}+(2^{n-k-1}-1)w_{k+2} \\
 \;\;\;\;\;\;\textrm{ if $k\equiv \floor{n/2}-1 $ or $\floor{n/2}-2 $ mod $4$}\\
w_k  \;\;\;\textrm{otherwise,}
\end{array} \right.
\end{displaymath}
with the only exception of
\begin{equation}\label{eq:projective y exceptions}
f_k({\bf w})
= \left\{ \begin{array}{ll}
w_{k-1}+(2^{n-k+1}-1)w_{k}+(2^{k-1}-1)w_{k-2} \\
 \;\;\;\;\;\;\textrm{ if $k={\frac{n}{2}}$ for $n$ even}\\
w_0+w_1(2^n-1) \\
\;\;\;\;\;\;\textrm{ if $k=0$ and $\floor{\frac{n}{2}}\not\equiv 1 \text{ or } 2\mod 4$ }
\end{array} \right.
\end{equation}

The idea is to write $f({\bf w})=\sum_{k=0}^nw_k\Gq{n}{k}$ as $\sum_{k=0}^ny_kf_k({\bf w})$, where $y_k$'s are some fixed non-negative real numbers. Note that ${\bf w}^*$ is the minimizer of all these cost functions in the feasible set and non-negativity of $y_k$'s automatically proves the optimality of ${\bf w}^*$ for $f({\bf w})$.

Given arbitrary $k$ such that $k\equiv \floor{\frac{n}{2}}+1\mod 4$ if non of the indices $\{k,k-1,k-2,k-3\}$ fall into the two exceptional categories in ~(\ref{eq:projective y exceptions}), we have,
\begin{align*}
&f_{k}({\bf w})\;\;\:\:\hspace{.12ex} =w_{k},\\
&f_{k-1}({\bf w})=w_{k-1},\\
&f_{k-2}({\bf w})=w_{k-2}(2^{k-1}-1)+w_{k-1}+w_k(2^{n-k+1}-1),\\
&f_{k-3}({\bf w})=w_{k-3}(2^{k-2}-1)+w_{k-2}+w_{k-1}(2^{n-k+2}-1).
\end{align*}

Furthermore, $\{w_{k},w_{k-1},w_{k-2},w_{k-3}\}$ show up only in these $f_i({\bf w})$'s. So, by comparing it to the corresponding coefficients in $f({\bf w})$, we must have
\begin{align*}
\Gq{n}{k}\;\;\;&=y_k+y_{k-2}(2^{n-k+1}-1)\\
\Gq{n}{k-1}&=y_{k-1}+y_{k-2}+y_{k-3}(2^{n-k+2}-1)\\
\Gq{n}{k-2}&=y_{k-2}(2^{k-1}-1)+y_{k-3}\\
\Gq{n}{k-3}&=y_{k-3}(2^{k-2}-1).
\end{align*}
By solving the system of equations above, we get
\begin{align*}\label{eq:projective y explicit}
y_{k-3}&=\Gq{n}{k-3}/(2^{k-2}-1)\geq 0\\
y_{k-2}&=(\Gq{n}{k-2}-y_{k-3})/(2^{k-1}-2)\geq 0\\
y_{k-1}&=\Gq{n}{k-1}-y_{k-2}-y_{k-3}(2^{n-k+2}-1)\geq 0\\
y_k&=\Gq{n}{k}-y_{k-2}(2^{n-k+1}-1)\geq 0.
\end{align*}
Story is not much different on the edges i.e. $y_0$ and $y_{\floor{\frac{n}{2}}}$ and we can prove the non-negativity of the first half of the $y_k$'s, which is followed by the non-negativity of the other half due to the symmetry. So $y_k$'s are non-negative and $w^*$ is the optimal transversal weight.

\begin{thebibliography}{10}
\providecommand{\url}[1]{#1}
\csname url@rmstyle\endcsname
\providecommand{\newblock}{\relax}
\providecommand{\bibinfo}[2]{#2}
\providecommand\BIBentrySTDinterwordspacing{\spaceskip=0pt\relax}
\providecommand\BIBentryALTinterwordstretchfactor{4}
\providecommand\BIBentryALTinterwordspacing{\spaceskip=\fontdimen2\font plus
\BIBentryALTinterwordstretchfactor\fontdimen3\font minus
  \fontdimen4\font\relax}
\providecommand\BIBforeignlanguage[2]{{%
\expandafter\ifx\csname l@#1\endcsname\relax
\typeout{** WARNING: IEEEtran.bst: No hyphenation pattern has been}%
\typeout{** loaded for the language `#1'. Using the pattern for}%
\typeout{** the default language instead.}%
\else
\language=\csname l@#1\endcsname
\fi
#2}}

\bibitem{BVP}
{C.\,Bachoc, F.\,Vallentin, and A.\,Passuello}, ``Bounds for projective codes from semidefinite programming,'' arXiv:1205.6406v2, Apr. 2013.

\bibitem{B79}
{C.\,Berge}, ``Packing problems and hypergraph theory: a Survey,'' \emph{Annals of Discrete Mathematics}, vol.\,4, pp.\,3--37, 1979.

\bibitem{BEV}
{M.\,Braun, T.\,Etzion, and A.\,Vardy}, ``Linearity and complements in projective space,'' \emph{Linear Algebra and its Applications}, vol.\,438, pp.\,57--70, Jan. 2013.

\bibitem{BYEB13}
{S.\,Buzaglo, E.\,Yaakobi, T.\,Etzion, and J.\,Bruck}, ``Error-correcting for multipermutations,'' \emph{IEEE Int. Symp. on Information Theory}, pp.\,724--728, Istanbul, Turkey, July 2013.

\bibitem{CSBB10}
{Y.\,Cassuto, M.\,Schwartz, V.\,Bohossian, and J.\,Bruck}, ``Codes for asymmetric limited-magnitude errors with application to multi-level flash memories,'' {\em IEEE Trans. Inform. Theory}, vol.\,56, no.\,4, pp.\,1582--1595, April 2010.

\bibitem{EB10}
{N.\,Elarief and B.\,Bose}, ``Optimal, systematic, $q$-ary codes correcting all asymmetric and symmetric errors of limited magnitude,'' {\em IEEE Trans. Inform. Theory}, vol.\,56, no.\,3, pp.\,979--983, March 2010.

\bibitem{TV11}
{T.\,Etzion and A.\,Vardy}, ``Error-correcting codes in projective space,'' \emph{IEEE Trans. on Information Theory}, vol.\,57, no.\,2, pp.\,1165--1173, Feb. 2011.

\bibitem{GYD13}
{R.\,Gabrys, E.\,Yaakobi, and L.\,Dolecek}, ``Correcting grain-errors in magnetic media,'' \emph{IEEE Int. Symp. on Information Theory}, pp.\,689--693, Istanbul, Turkey, July 2013.

\bibitem{GYD13b}
{R.\,Gabrys, E.\,Yaakobi, and L.\,Dolecek}, ``Correcting grain-errors in magnetic media,'' submitted to \emph{IEEE Trans. on Information Theory}, 2013.

\bibitem{G52}
{E.N.\,Gilbert}, ``A comparison of signalling alphabets," \emph{Bell System Technical Journal}, vol.\,31, pp.\,504--522, 1952.

\bibitem{ISW11}
{A.R.\,Iyengar, P.H.\,Siegel, and J.K.\,Wolf}, ``Write channel model for bit- patterned media recording,'' \textit{IEEE Transactions on Magnetics}, vol.\, 47, no.\,1, pp.\,35--45, Jan. 2011.

\bibitem{KZ13}
{N.\,Kashyap and G.\,Z\'{e}mor}, ``Upper bounds on the size of grain-correcting codes,'' arXiv:1302.6154v4, Jun. 2013.

\bibitem{KBE10}
{T.\,Kl{\o}ve, B.\,Bose, and N.\,Elarief}, ``Systematic single limited magnitude asymmetric error correcting codes,'' {\em Proc. IEEE Inform. Theory Workshop}, Cairo, Egypt, January 2010.

\bibitem{KK08}
{R.\,Koetter and F.R.\,Kschischang}, ``Coding for errors and erasures in random network coding,'' \emph{IEEE Trans. on Information Theory}, vol.\,54, no.\,8, pp.\,3579--3591, Aug. 2008.

\bibitem{KK12}
{A.\,Kulkarni and N.\,Kiyavash}, ``Non-asymptotic upper bounds for deletion correcting codes,'' arXiv:1211.3128v1, November 2012.

\bibitem{MBK11}
{A.\,Mazumdar, A.\,Barg, and N.\,Kashyap}, ``Coding for high-density recording on a 1-d granular magnetic medium,'' \emph{IEEE Transactions on Information Theory}, vol.\,57, no.\,11, pp.\,7403--7417, Nov. 2011.

\bibitem{S14}
{M.\,Schwartz}, ``On the non-existence of lattice tilings by quasi-crosses,'' \emph{European J. of Combinatorics}, vol.\,36, pp.\,130--142, Feb. 2014.

\bibitem{S62}
{F.F.\,Sellers, Jr}, ``Bit loss and gain correcting code,'' \emph{IRE Trans. on Information Theory}, vol.\,8, no.\,1, pp.\,35--38, January 1962.

\bibitem{SR11}
{A.\,Sharov and R.M.\,Roth}, ``Bounds and constructions for granular media coding,'' \emph{IEEE Int. Symp. on Information Theory}, pp.\,2343--2347, St. Petersburg, Russia, August 2011.

\bibitem{SR13}
{A.\,Sharov and R.M.\,Roth}, ``Improved bounds and constructions for granular media coding,'' \emph{Proc. Allerton Conf. Commun., Control and Comput.}, Allerton Retreat Center, Monticello, IL, 2013.

\bibitem{T97}
{J.M.G.M.\,Tolhuizen}, ``The generalized Gilbert-Varshamov bound is implied by Tur\'{a}n's theorem,'' \emph{IEEE Trans. on Information Theory}, vol.\,43, pp.\,1605--1606, September 1997.

\bibitem{V57}
{R.R.\,Varshamov}, ``Estimate of the number of signals in error correcting codes,'' \emph{Dokl. Acad. Nauk SSSR}, vol.\,117, pp.\,739--741, 1957.

\bibitem{VT65}
{R.R.\,Varshamov and G.M.\,Tenengolts}, ``Codes which correct single asymmetric errors (in Russian),'' \emph{Automatika i Telemekhanika}, vol.\,26, np.\,2, pp.\,288--292, 1965. English translation in \emph{Automation and Remote Control}, vol.\,26, no.\,2, pp.\,286--290, 1965.

\bibitem{WVB88}
{J.\,Weber, C.\,De Vroedt, and D.\,Boekee}, ``Bounds and constructions for binary codes of length less than 24 and asymmetric distance less than 6,''
\emph{IEEE Trans. on Information Theory}, vol.\,34, no.\,5, pp.\,1321--1331, September 1988.

\bibitem{WVB87}
{J.\,Weber, C.\,De Vroedt, and D.\,Boekee}, ``New upper bounds on the size of codes correcting asymmetric errors,''
\emph{IEEE Trans. on Information Theory}, vol.\,33, pp.\,434--437, May 1987.

\bibitem{WNP97}
{R.I.\,White, R.M.H.\,New, and R.F.W.\,Pease}, ``Patterned media: viable route to 50Gb/in2 and up for magnetic recording,'' \textit{IEEE Trans. Magnetics}, vol.\,33, no.\,1, pt.\,2, pp.\,990--995, Jan. 1997.

\bibitem{WWKM09}
{R.\,Wood, M.\,Williams, A.\,Kavcic, and J.\,Miles}, ``The feasibility of magnetic recording at 10 terabits per square inch on conventional media,'' \textit{IEEE Transactions on Magnetics}, vol.\,45, no.\,2, pp.\,917--923, 2009.

\bibitem{YMGSSW10}
{E.\,Yaakobi, J.\,Ma, L.\,Grupp, P.H.\,Siegel, S.\,Swanson, and J.K.\,Wolf}, ``Error characterization and coding schemes for flash memories,'' \emph{Proc. Workshop on the Application of Communication Theory to Emerging Memory Technologies}, Miami, Florida, December 2010.
\end{thebibliography}
\end{document}